\documentclass[10pt,twocolumn,letterpaper]{article}

\usepackage{times}
\usepackage{epsfig}
\usepackage{graphicx}
\usepackage{amsmath}
\usepackage{amsthm,amsfonts,amssymb}
\usepackage{subfig}

\def\mat#1{\mathchoice{\mbox{\boldmath$\displaystyle\tt#1$}}
{\mbox{\boldmath$\textstyle\tt#1$}}
{\mbox{\boldmath$\scriptstyle\tt#1$}}
{\mbox{\boldmath$\scriptscriptstyle\tt#1$}}}

\def\vec#1{\mathchoice{\mbox{\boldmath  $\displaystyle\bf#1$}}
{\mbox{\boldmath  $\textstyle\bf#1$}}
{\mbox{\boldmath  $\scriptstyle\bf#1$}}
{\mbox{\boldmath  $\scriptscriptstyle\bf#1$}}}

\newlength{\colwidth}
\newcommand{\SKIP}[1]{} 

\newcommand{\mbegin} {\left [ \begin{array}}

\newcommand{\mend}   {\end{array} \right ]}

\newcommand{\vbegin} {\left ( \begin{array}{c}}

\newcommand{\vend} {\end{array}\right )}

\def\squareforqed{\hbox{\rlap{$\sqcap$}$\sqcup$}}

\def\qed{\ifmmode\squareforqed\else{\unskip\nobreak\hfil
	\penalty50\hskip1em\null\nobreak\hfil\squareforqed
	\parfillskip=0pt\finalhyphendemerits=0\endgraf}\fi}

\def\vec#1{\mathchoice%
	{\mbox{\boldmath $\displaystyle\bf#1$}}
	{\mbox{\boldmath $\textstyle\bf#1$}}
	{\mbox{\boldmath $\scriptstyle\bf#1$}}
	{\mbox{\boldmath $\scriptscriptstyle\bf#1$}}}

\newcommand{\showeqnlabel}{
	\hbox to 0pt{\quad\quad\relax\fbox{\scriptsize\rm\eqnlblx}%
	\gdef\eqnlblx{xxxx}}} \newcommand{\eqnlblx}{}

\def\@eqnnum{\rm (\theequation)\showeqnlabel}

\newcommand{\nofig}[1]{\centerline{\bf Figure here}}

\def\mat#1{\mathchoice{\mbox{\boldmath$\displaystyle\tt#1$}}
	{\mbox{\boldmath$\textstyle\tt#1$}}
	{\mbox{\boldmath$\scriptstyle\tt#1$}}
	{\mbox{\boldmath$\scriptscriptstyle\tt#1$}}}

\usepackage{algorithm}
\usepackage{algorithmic}

\begin{document}

\title{Uncalibrated 3D Room Reconstruction from Sound}

\author{Marco Crocco$^*$, 
        Andrea Trucco$^{*+}$,
        and~Alessio~{Del Bue}$^*$\\
        \\
        	$^*$Pattern Analysis and Computer Vision Department (PAVIS),\\
        	Istituto Italiano di Tecnologia
        	Via Morego 30, 16163 Genova, Italy\\
        	$^+$ Department of Electrical, Electronic, Telecommunications Engineering, \\and Naval Architecture (DITEN), University of Genoa, 16145 Genova, Italy} 
%

\maketitle
\begin{abstract}

This paper presents a method to reconstruct the 3D structure of generic convex rooms from sound signals. Differently from most of the previous approaches, the method is fully uncalibrated in the sense that no knowledge about the microphones and sources position is needed. Moreover, we demonstrate that it is possible to bypass the well known echo labeling problem, allowing to reconstruct the room shape in a reasonable computation time  without the need of additional hypotheses on the echoes order of arrival. Finally, the method is intrinsically robust to outliers and missing data in the echoes detection, allowing to work also in low SNR conditions. The proposed pipeline formalises the problem in different steps such as time of arrival estimation, microphones and sources localization and walls estimation. After providing a solution to these different problems we present a global optimization approach that links together all the problems in a single optimization function. The accuracy and robustness of the method is assessed on a wide set of simulated setups and in a challenging real scenario. Moreover we make freely available for a challenging dataset for 3D room reconstruction with accurate ground truth in a real scenario.      
\end{abstract}

\section{Introduction}

Sensing the shape of a room using acoustic signals is a problem that has attracted increasing attention from the research community. This is mainly because room reconstruction  is an enabling technology for the ubiquitous localisation of ``things'' in indoor environments with the simple use of inexpensive sensors such as the microphones in mobile phones and other consumer products. Such interest has however clashed against the complexity of the task of localising the position of the walls, sensors and sources by analysing only a set of acoustic events. These being, in the most blind scenario, unknown signals generated from unknown sources with an arbitrary position. Such attractiveness has although clashed against the intrinsic complexity of the problem. Current solutions often need custom hardware requirements or even constraints that make the applicability of each method subject to the specific setup or to limiting assumptions. 

On the contrary, we are dealing with the general optimization problem for room reconstruction where each source generates a sound event (not impulsive) which is acquired by a set of microphones deployed randomly in an unknown indoor area. The solution of such optimization is the 3D metric positions of the microphones, sources and the room wall positions. Crucially, the cost function derived from this problem is non-convex, highly non-linear and with several ambiguous solutions. Moreover, the solution in such general form corresponds to solve for three different problems that formally have been treated separately: microphone positions estimation, source localization and room reconstruction. For this reason, most of the approaches were devised to solve, or to consider solved, a subset of these three problems. 
A common assumption of many approaches is knowing a priori the  positions of the microphones and/or the acoustic sources \cite{Filos2010,Tervo2012,Ribeiro2012,Antonacci2010,Dokmanic2013,Tervo2010}. In addition, many methods require specific microphone arrays or source arrangements to avoid ambiguities in the order of arrival of the reflections \cite{Ribeiro2012,Antonacci2010,Tervo2012}.  
%
Differently, this paper shows that it is possible to obtain a solution even if 
the resulting cost function is strongly non-linear and characterized by a high number of variables.  

\begin{figure*}[thb]
	\begin{center}
		\includegraphics[width=.8\textwidth]{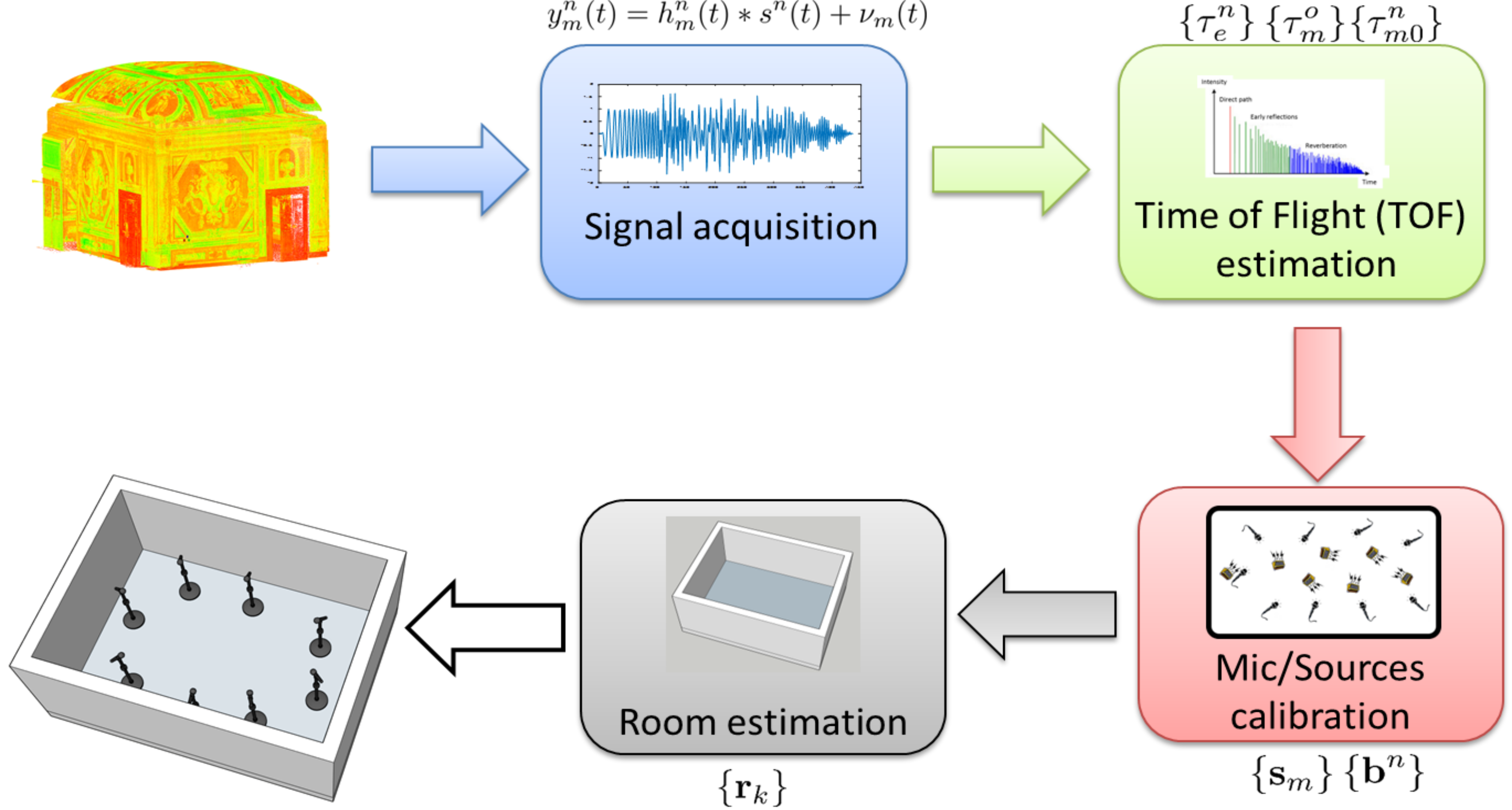}
		\caption{The scheme presenting the pipeline for 3D room reconstruction.}
		\label{fig:pipeline_scheme}
	\end{center}
\end{figure*}

The devised strategy can be summarised in Figure \ref{fig:pipeline_scheme}:  
Given a set of microphones deployed in a room and a set of sources or a moving source, both with unknown positions, a set of signals are acquired by the microphones. Given such recordings of the sound events, Times of Flight (TOFs) for the direct path and reflections are estimated by compressing the signals with a matched filter thus extracting the resulting peaks. Since first peaks in order of arrival correspond to direct paths, they are not subject to ambiguities and can be exploited to estimate a first guess of microphones and sources positions by an iterative procedure \cite{Gaubitch2013} having at the core a matrix factorization-based solution \cite{Crocco2012}.  Once microphones and sources positions are available, planar surface positions can be estimated by an exhaustive search over a discretized grid in the space of all possible configurations, evaluating a proper cost function parametrized by microphone and source positions and signal peaks. In this way, the intrinsic ambiguity problem arising from the unknown matching between signal peaks and planar surfaces (the well known echo-labeling or echo-sorting problem) is bypassed. However, as the 3D space of configurations is of dimension three multiplied by the number of planar surfaces, a naive exhaustive search would be computationally infeasible. For this reason we adopt a greedy iterative procedure, the core of our approach, in which search is decoupled for each planar surface, decreasing the space of solution to 3 for each wall search. To do this, we initially discard second order reflections since they imply dependencies between different planar surfaces even if we reconsider them as long as newly estimated surfaces are available. Moreover, at each iteration, we prune out peaks matched to an already estimated planar surface to simplify the search over subsequent surfaces. Finally we adopt a robust cost function that allows to cope with missing peaks or spurious peaks due to not perfect functioning of the peak finder stage. The results of this stage is a collection of first guess planar surfaces positions together with a labelling that links peaks in the signal to surfaces. These data are finally used to build a cost function, whose variables are microphones, sources, surfaces positions and onset times, which is continuous and differentiable since the matching between walls and delays is already solved. This allows to perform global optimization with nonlinear Least Squares further refining the solution. 

\section{Related work}

The room estimation problem is a complex procedure that can be divided in different sub-problems each one related to a particular field in Signal Processing. Of course there are works that deal entirely with the full problem \cite{Dokmanic2013,dokmanic2014localize,Crocco:DelBue:2014,Ribeiro2012,Antonacci2012,Antonacci2010,Filos2010,Filos2011,Wang2016,Remaggi2014,Remaggi2015,Tervo2010,Rajapaksha2016,Tervo2012}, from signals to 3D room estimation, but still it would be reductive to consider only those. Moreover each step and relative methods in room reconstruction have their own assumptions that reduce their applicability; thus a careful review of the state of the art at each stage is necessary.

\subsection{Time of arrival estimation}

For the most part of methods \cite{Antonacci2010,Antonacci2012,Filos2010,Filos2011,Tervo2012,Dokmanic2013,dokmanic2014localize,Remaggi2014,Remaggi2015,Remaggi2015_mics,Wang2016,Rajapaksha2016} the initial step is to move from signals to distances through estimation of time of arrival (TOA) of the direct path and echoes of the signal from the source (either real or virtual) to the microphone. This step is necessary to estimate geometry since distances can be used to recover back 3D positions and walls orientations. 
In most cases the emitted signal is known so TOAs are estimated by detecting the relevant peaks in the cross-correlations between the emitted signal and the signals received at the microphones. To improve the estimation accuracy generalized cross correlation 
\cite{Knapp1976} is sometimes adopted. 
Concerning the peak-finding step, among the software tools available (e.g. Voicebox\footnote{{www.ee.ic.ac.uk/hp/staff/dmb/voicebox/voicebox.html}.} ), there are solutions tuned to signal with specific statistics like speech \cite{brookes2006quantitative,Drugman:etal:2012}. These algorithms perform local maxima estimation pruning out peaks due to noise or signal side-lobes given by limited bandwidth. 
If the transmitted signal is unknown cross-correlation can be computed between couples of received signals at different microphones, but in this case spurious peaks arise due to the interaction between the echoes corresponding to different reflectors. An alternative non-linear optimization method \cite{Crocco:DelBue:2014} jointly estimates transmitted signal and TOAs. To cope with the sensitivity to initialization and non-convexity of the problem Simulated Annealing is employed resulting in relevant computational costs.

Finally, it is worth mentioning a set of blind method that relies on the intrinsic sparsity of the first part of the room impulse response (RIR) \cite{Kowalczyk2013}. More recent approaches also include non-negative constraints to increase the accuracy of the solution \cite{Lin2007NNeg,Crocco:DelBue:2015,Crocco:DelBue:2016}. These methods firstly estimate the RIR without the knowledge of the transmitted signal using an iterative sequence of convex problems. Then, the peaks of the RIR  provide the TOAs related to the room.

It is crucial to note that, for every TOA estimator,  a practical trade off exists between the number of missed TOAs and the number of spurious TOAs wrongly selected. This trade-off is only partially dependent by the SNR since, also for clean received signals, many factors can provide spurious peaks. For instance, side lobes due to finite signal bandwidth, echo distortions due to frequency dependent attenuations and coalescing peaks due to close TOAs can affect peak estimation. This fact is often a source of unavoidable outliers that make the robustness of subsequent steps in room estimation a delicate and very important issue.

Another issue characterizing all the methods is that the echo labeling problem, i.e. the fact that apart for the first TOA that is linked to the direct path, the association between TOAs and reflecting surfaces is not known, so giving rise, in absence of further a priori knowledge assumed by many methods, to a NP-hard combinatorial problem.  

\subsection{Microphones and sound events localisation}

Given the estimated TOAs for the different sources at the microphones, it is possible to measure distances between them by knowing time of emission (TOE) of the sound events and the sampling offset time at each microphone. These two quantities can be assumed known if the acquisition system accounts for a shared clock between all microphones and sources. But in the most general scenario these timings are unknown and they have to be estimated from the TOAs in order to obtain the exact time of flights (TOFs). Pollefeys and Nister \cite{Pollefeys:Nister:2008} first presented a closed-form solution for estimating TOFs from TOAs given a negligible sampling offset time. Most notably, the method shows that if $10$ microphones and $5$ sources are available, it is possible to formulate the time of emission estimation as a linear Least Squares estimation problem. This however requires a grouping of microphones/sources in sets of 10/5 respectively that might lead to sub-optimal solutions. Differently, the work in \cite{Jiang:etal:2013} shows that by rearranging the TOAs in a matrix, if the TOEs are correctly estimated the matrix becomes rank-constrained. This constraint is used to instantiate an optimization problem over the TOEs minimizing the nuclear norm of the TOFs matrix. Similarly, Gaubitch et al. \cite{Gaubitch2013} showed that both TOEs and sampling offsets can be estimated with an iterative algorithm from a minimum configuration of $5$ microphones and $13$ sources.

Once TOFs are known, squared distances can be computed by knowing the sound speed coefficient $c$. These distances can be re-arranged into a matrix $\mat D$ that in the noiseless class is rank constrained (i.e. $rank(\mat D) \leq 5$). This property was first made explicit by Thrun \cite{thrun2005affine} for far field conditions. Interestingly the rank of $\mat D$ make possible to extract microphones and sources positions without any a priori knowledge since such constraint encodes the quadratic relations that gives distances in terms of microphone and sound sources position.  Crocco et al. \cite{Crocco:etal:2012} made explicit this relation with a bilinear form and proposed a closed-form solution to extract the 3D positions if one source coincides with one microphone. Dokmanic et al. \cite{dokmanic2015relax} instead arranged distances with a square Euclidean Distance Matrix (EDM) and exploited the properties of such matrix to solve for the localisation of microphones/sources. Also in \cite{dokmanic2014localize}, the fact that the sound propagates in a room can be used to calibrate 3D positions by consider additional sources given by the echos of the walls' reflections. More recently, Le and Ono \cite{TRUNG2016} provided a very appealing closed-form solution for the case of $9$ sources and $4$ sensors starting from the distance matrix used in \cite{Crocco:etal:2012}. 

It is also worth noting that just a few room reconstruction methods assume both microphone and source positions as unknown \cite{Crocco:DelBue:2014,dokmanic2014localize,Remaggi2015_mics}, while a significant number of methods  \cite{Antonacci2012,Remaggi2015,Remaggi2014,Filos2010,Dokmanic2013} starts from a known and often not arbitrary microphone displacement (i.e. a compact microphone array) and infers the real source location by triangulation \cite{Antonacci2012,Filos2010} or a combination of distance and direction of arrival estimation \cite{Remaggi2015,Remaggi2014}. 

\subsection{Room geometry estimation}

The estimation of reflectors position and orientation is a problem solved with two different approaches: \textbf{Direct methods} using only the signals and \textbf{two steps methods} starting from the TOFs to reach the estimation of the room geometry.

\subsubsection{Direct Methods}

This class of methods use the signal at the microphones to directly estimate walls positions thus avoiding the peak detection step. 

The work of Tervo and Korhonen \cite{Tervo2010} performs a brute force search to find the maximum of a cost function based on crosscorrelations among all microphone couples. The approach estimates the reflector positions whose computed TOFs best match the maxima of all the cross-correlations forming the cost function.  The appeal of the method is strongly limited by the fact that it is devised for just one significant planar reflector. This limitation obviously eliminates the problems of echo labeling and of the presence of spurious maxima in the cross-correlation due to the interaction between echoes from different reflectors. 

Ribeiro et al. \cite{Ribeiro2012} propose an approach based on sparse modelling using a dictionary encoding the different walls positions. Each word of the dictionary encodes the RIR associated to a reflector position and it can be selected by optimizing a cost function with a sparsity promoting term. The construction of the dictionary is made by simulating or measuring the whole set of possible RIRs -- a time consuming step that requires the information of the microphone positions (in a compact spatial configuration). Further hypothesis for the method to provide a solution is the assumption of floor and ceiling being the strongest reflectors in the room. Finally, the sparsity is imposed by l1 regularization, a convex relaxation of l0 regularization that is not guaranteed to provide a sparse solution for every possible condition.

\subsubsection{Two-step approaches}

Two-step approaches assume that TOAs ot TOFs are already known and attempt to use this information to estimate the reflectors.

In a line of research \cite{Antonacci2010,Filos2010,Filos2011,Antonacci2012,Remaggi2014,Remaggi2015,Remaggi2015_mics} reflectors are modelled as planes tangent to the ellispoids with foci given by each pair of microphone/source.   The condition of tangency of multiple ellipses is not assured in the real case because of noise. Thus, non-linear optimization, Hough transform or RANSAC based approaches are used in order to provide a robust solution for each reflector. The method has been first applied to the 2D case \cite{Antonacci2010,Antonacci2012,Filos2010,Filos2011}, assuming perfectly absorbing floor and ceiling and then extended to generic 3D rooms \cite{Remaggi2014,Remaggi2015,Remaggi2015_mics}. The amount of a priori knowledge was gradually decreased from the first works requiring both microphone and source position knowledge \cite{Antonacci2010,Filos2011} passing through extensions requiring just the knowledge of microphone positions \cite{Antonacci2012,Filos2010,Remaggi2014,Remaggi2015} and arriving to completely uncalibrated setups where both microphone and source positions are estimated \cite{Remaggi2015}. However in the latter case \cite{Remaggi2015} a rough initialization by hand of microphones positions is still required.  

The echo sorting problem is faced in \cite{Antonacci2010} using a clustering procedure of TOFs based on the Hough transform; however this method requires a very specific setup with just one microphone and a source moving on a perfect circle around the microphone.  Differently in \cite{Antonacci2012}, it is assumed that the source is moved very close to a reflector at each acquisition, such that the second TOF is surely related to the same reflector for all the microphones. In \cite{Filos2010} a brute force search is run over all the possible TOF combinations and once a reflector is estimated the corresponding TOFs are iteratively discarded: The approach becomes feasible given the low number of microphones/sources and by the fact that just first order TOFs are assumed to be detected. The rest of the approaches  \cite{Filos2011,Remaggi2014,Remaggi2015,Remaggi2015_mics} does not provide specific solutions for the echo sorting problems. 

Despite the relative robustness given by Hough transform or RANSAC, the above methods do not have specific strategies to cope with missing or spurious TOFs given by malfunctioning of the peak finder or by selection of peaks corresponding to high order reflections\footnote{In some cases it is explicitly stated that such events are discarded in the evaluation of results.}


 
 A different approach \cite{Dokmanic2013} uses the EDM matrix initially constructed by known microphones positions and augments such matrix iteratively picking a subset TOFs, one from each microphone. If the augmented matrix is again close to an EDM, the subset of TOFs is related to the same virtual source and the subset is used to estimate it and subsequently removed from the whole set of TOFs. After pruning out the virtual sources  corresponding to higher order echoes the whole geometry is estimated.  
 The method alleviates the echo labelling problem since the number of TOFs to test is gradually lowered with the iterations. However for a moderate number of microphones the number of TOFs subsets to test is still very high and some heuristics based on the microphone array compactness are required. Moreover the method is not robust to missing TOFs since the EDM based criteria proposed to select the TOF subsets are very sensitive to outliers. 
 The EDM criterion is exploited also in \cite{dokmanic2014localize} in which  microphones and virtual source positions are jointly estimated by using just one real source location, thus providing a complete reconstruction in a fully uncalibrated way. Unfortunately, the echo labelling problem and the lack of robustness to outliers is stronger than in \cite{Dokmanic2013} since no initial EDM matrix given by inter-microphone distances is available.
 
 A maximum likelihood approach is devised by \cite{Tervo2012} where a nonlinear cost function based on the differences between measured TOFs and TOFs computed from the guessed virtual sources positions is minimized. 
 The reflectors positions are then geometrically estimated from the virtual source positions. A very strong simplifying assumption, i.e. the order of arrival of echoes is the same for all microphones, allows to avoid the echo labelling problem, but it holds only for very compact microphones. Moreover the maximum likelihood cost function is based on a Gaussian noise assumption that does not consider outliers due to outliers in peaks detection. 
 
 
 Instead, the approach of Crocco and Del Bue \cite{Crocco:DelBue:2014} define the room geometry estimation problem as an optimization problem without any a priori information (apart from the room convexity assumption). The method firstly estimates the times of arrival and the transmitted signal exploiting an hybrid strategy, partly analytic and partly stochastic, using simulated annealing. Then, peaks corresponding to direct path arrivals are used to estimate a first guess micropone and source positions by a bilinear approach. Finally, given the whole set of extracted peaks, simulated annealing is again employed to estimate the planar reflectors positions and to refine the microphone and source localisation. In addition, a pruning strategy is devised to discard ambiguous peaks during optimisation. 
 
 Among the most recent works, \cite{Wang2016} recovers the 2D room geometry by just a single mobile device moved in different locations, providing that the microphones and speakers are co-located in the device and that the distance between subsequent locations can be measured. The work has a number of limitations, including the need of heuristics on the angle between adjacent walls, the poor performance when high order TOFs are wrongly classified as first order ones and the need to perform a combinatorial search over all the possible echo combinations. 
 
 Another recent work \cite{Rajapaksha2016} estimates all the real and virtual source positions by a closed form approach; from each virtual source a tentative reflector position is inferred and finally reflectors corresponding to high order virtual sources are pruned out by a geometric reasoning. The method has the merit of solving the issue of high order echoes but it does not deal with the echo labelling problem, requiring an exhaustive combinatorial search. Moreover the first step based on closed form solution is not robust to missing data: a missing first order TOF could completely break out the algorithm. 


\section{Problem statement}
\begin{figure}[h]
	\begin{center}
		\includegraphics[width=1\linewidth]{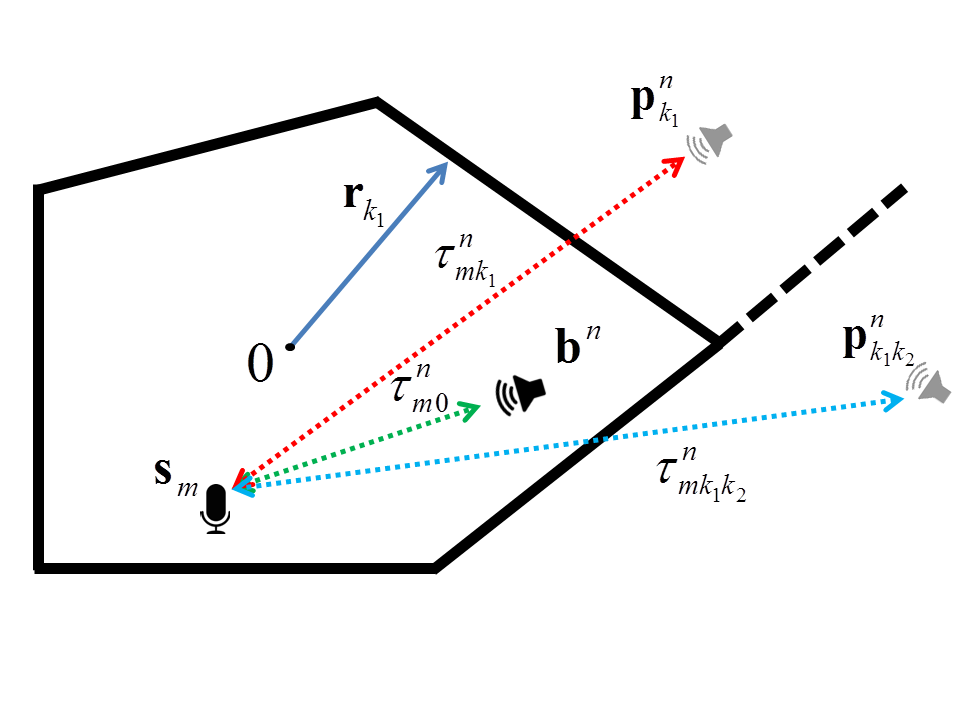}   
	\end{center}
	\vspace{-.35cm}
	\caption{Schematic view of room, microphone, real source, first and second order virtual sources and times of flight. }
	\label{fig:room_scheme}
	\vspace{-.2cm}
\end{figure}

Let us consider $N$ point-like, omnidirectional audio sources displaced in a 3D environment, whose 3D positions are given by the 3-vectors $\vec b^n$ with $n=1,\ldots,N$. Similarly, consider $M$ point-like omnidirectional microphones whose 3D positions are given by the 3-vectors $\vec s_m$ with $m=1,\ldots, M$. Sources and microphones are enclosed in a room whose boundaries define a convex polyhedron with a known number of faces $K$. The 3D positions of these $K$ planar sufaces are defined by the 3-vectors $\vec r_k$ with $k=1,\ldots, K $, normal to the planar surfaces and with modulus equal to the distance of each planar surface from the 3D coordinate center (see Fig. \ref{fig:room_scheme}).  

Each source emits a signal $x^n(t)$, in general different from source to source.  The Time of Flight (TOF) $\tau^n_{m0}$ between a source $n$ and a microphone $m$, can be expressed as the microphone-source distance divided by the sound velocity $c$, as follows:
\begin{equation}
\tau^n_{m0}=\frac{\|\vec b^n-\vec s_m\|_2}{c}.
\label{i1} \end{equation}
TOF is the time needed by a signal to propagate from a source to a microphone along the direct path, i.e. in free space conditions.
However, in an enclosure the propagating signal will be reflected by the boundaries, and the signal received at each microphone will be the sum of several contributions related to the multiple reflections occuring in the room. The relation between a source emitted at $\vec b^n$ and the signal acquired at $\vec s_m$ is modeled by the Room Impulse Response (RIR), defined by the signal arriving at point $\vec s_m$ when a Dirac pulse is emitted at $\vec b^n$. In a general case it is very difficult to reliably model the RIR; however when the room is a convex polyhedron and the wavelengths involved in the transmitted signal are small in comparison to the areas of the planar surfaces, the first portion of the RIR can be fairly approximated with the image model \cite{Allen1979}. According to this model, when a source $n$ emits a signal, the corresponding reflection from the surface $k$ is modelled as a virtual source emitting the same signal synchronously with the real one, whose 3D position $\vec p^n_k$ is symmetric to the real source position, taking the surface plane as the symmetry plane (see Fig. \ref{fig:room_scheme}):
\begin{equation}
\vec p^n_k  = IM(\vec b^n,\vec r_k) = \vec b^n+2\left( 1- \frac{ \vec r^{\top}_k \vec b^n}{\|\vec r_k\|^2_2}  \right)\vec r_k,
\label{i2}
\end{equation}
where $IM(\vec b^n,\vec r_k)$ is defined as the function producing the specular image of $\vec b^n$ with respect to the plane $\vec r_k$.
In practice the planar surface is considered to be a perfect acoustic mirror and the virtual source is the acoustic image of the real one.
In general the signal reflected from a surface impinges on other surfaces, yielding high order reflections. Here we take into account just second order reflections that can be again reliably modeled with the image method. In particular $K(K-1)$ second order virtual sources can be identified, corresponding to all the ordered couples of different $K$ surfaces. Their location $\vec p^n_{k_1k_2}$ for $k_1,k_2 = 1, \ldots, K$ and $k_1 \neq k_2$ is given by:
\begin{equation}
\vec p^n_{k_1k_2} = IM(\vec p^n_{k_1},\vec r_{k_2}) =\vec p^n_{k_1}+2\left( 1- \frac{ \vec r^{\top}_{k_2} \vec p^n_{k_1}}{\|\vec r_{k_2}\|^2_2}  \right)\vec r_{k_2}.
\label{eq:image}
\end{equation}
The TOF at microphone $m$ related to a virtual source of first or second order is given by:
\begin{equation}
\tau^n_{mk} = \frac{\|\vec p^n_k - \vec s_m\|_2}{c}, \hspace{0,5cm} \tau^n_{mk_1k_2} = \frac{\|\vec p^n_{k_1k_2} - \vec s_m\|_2}{c}
\label{dist}.
\end{equation} 
According to the image model, the signal at a microphone location can be approximated with a sum of scaled and delayed replicas of the transmitted signal, each one corresponding to a direct path, first or second order reflection. We neglect frequency dependent attenuation occurring during propagation, nonlinear effects and frequency dependent reflections.

The time of arrival (TOA) of each signal component, i.e. the time at which a signal component is detected at a microphone depends from three terms: the TOF, the time of emission from the $n$-th real source $\tau^n_e$, and the offset time $\tau^o_m$ related to the clock reference time at microphone $m$. Thus, denoting the TOAs related to direct path, first and second order image sources with $\tau^{an}_{m0}$, $\tau^{an}_{mk}$ and $\tau^{an}_{mk_1k_2}$, respectively we have: 
\begin{equation}
\tau^{an}_{m0}=  \tau^{n}_{m0}+\tau^n_e+\tau^o_m, 
\end{equation}
\begin{equation}
\tau^{an}_{mk}=  \tau^{n}_{mk}+\tau^n_e+\tau^o_m,
\end{equation}
\begin{equation}
\tau^{an}_{mk_1k_2}=  \tau^{n}_{mk_1k_2}+\tau^n_e+\tau^o_m. 
\end{equation}
Apart from the offset times, we assume that all the microphones have the same impulse response. Likewise, if a loudspeaker is employed to generate the sources, we can assume the same impulse response for each source. Thus we can include both impulse responses in the transmitted signal by defining:
\begin{equation}
s^n(t) = h^{mic}(t)*h^{speaker}(t)*x^n(t).
\end{equation}
Given the above definition, we can express the relationship between signal $y^n_m(t)$ acquired at a microphone $m$ when the $n$-th real source emits, and the corresponding signal $s^n(t)$, as 
%
\begin{eqnarray}
y^n_m(t)=(\sum_{k=0}^K  a^n_{mk}\delta(t-\tau^{an}_{mk})+ \nonumber \\ +\sum_{k_1,k_2=1}^K a^n_{mk_1k_2}\delta(t-\tau^{an}_{mk_1k_2})+ \nonumber \\ +z^n_m(t))*s^n(t)
\end{eqnarray}
where $\delta(t)$ denotes the Dirac pulse, $a^n_{mk}$ 
are the amplitude coefficients related to direct path, first and second order reflections and $z^n_m(t)$ is a component encompassing higher order reflections, border effects, possible reflections from objects in the room, etc. Notice that we assume the amplitude coefficients to be frequency-independent.

After formalising the involved equations, now we can formulate the problem we want to address. Given the set of signals $y^n_m(t)$ with $m=1,\ldots,M$ and $n=1,\ldots,N$, recover the 3D positions of microphones $\vec s_m$, sources $\vec b^n$ and planar surfaces $\vec r_k$. Notice that the only prior knowledge is the number of planar surfaces and the room convexity, while we do not make any assumption on the relative displacement of microphones and sources. 

\section{Proposed method}

\subsection{From signals to times of arrival}
As transmitted source we use a frequency sweep pulse (chirp), defined as:
\begin{equation}
x(t)=\sin(2\pi f_0t+f_1t^2), 
\end{equation}
 where $f_0$ and $f_1$ are the initial and final instantaneous frequencies and $t \in [0, (f_1-f_0)/2f_1]$. The chirp pulse is emitted by a loudspeaker moved in different positions. To recover the temporal resolution needed to resolve the signal components, the signals acquired by the microphones are filtered with a matched filter $m(t)$. The latter is defined by a time reversed version of the transmitted signal, whose spectrum has been normalized by the spectrum modulus, according to the generalized cross/correlation method \cite{Knapp1976}. 
The final compressed signal for microphone $m$ and source $n$ is obtained as: 
\begin{equation}
\hat{y}^n_m(t) = y^n_m(t)*m(t).
\end{equation}
The compressed signal exhibits a set of peaks related to each signal component, and the time at which a peak occurs is equal to the TOA of the corresponding component. In order to robustly detect peaks and evaluate the corresponding TOAs we adopt the procedure as described in Algorithm 1.
\begin{center}
\begin{algorithm}
	\caption{Procedure to detect relevant peaks in the acquired signal.} 
	\label{alg:Peaks}
	\begin{algorithmic}[1]
		\STATE set $thr_1$ and $thr_2$
		\FOR {$t = t^{start}$ to $t^{end}$}
		\STATE find the local maximum $lm(t)$ of the signal in a sliding window of size $T$:
		$lm(t) =\max \left\{\hat{y}^n_m(t),\ldots,\hat{y}^n_m(t+T)\right\}$
		\STATE set to zero all the values in the window lower than $thr_1 \cdot lm(t)$:
		\ENDFOR 
		\STATE find the global maximum $gm$ of the signal $\hat{y}^n_m(t)$. 
		\STATE  set to zero all the values of $\hat{y}^n_m(t)$ lower than $thr_2 \cdot gm$.
		\STATE $l = 1$ 
		\FOR {$t = t^{start}$ to  $t^{end}$}
		\IF {$\hat{y}^n_m(t)\neq 0$}
		\STATE $\tilde{\tau}^{a,n}_{m,l}= t$
		\STATE  $l = l+1$
		\ENDIF
		\ENDFOR
	\end{algorithmic}
\end{algorithm}
\end{center}
%
The algorithm begins with a thresholding operation over the sliding window that results in a non maxima suppression that discards side lobe peaks of each signal replica. To this end, the sliding window length is set roughly equal to the compressed pulse length and a first threshold $thr_1$ is set to a value close to one, in order to discard almost all the samples belonging to a pulse and different from its maximum, and, at the same time preserving peaks corresponding to other pulses that could be very close to the current one. The second global thresholding operation over the whole signal discards peaks likely due to noise. Consequently, the threshold $thr_2$ is set according to the estimated SNR of the received signals, in order to empirically maximize the probability of discarding noise while minimizing the probability of discarding reflections of small intensity. 
The above algorithm is repeated for all the microphone-source couples, collecting the estimated TOAs $\tilde{\tau}^{a,n}_{m,l}$.  Notice that, fixed $n$ and $m$, the number of TOAs per received signal is in general different, i.e. $l = 1,\ldots,L(n,m)$, 
due to missing data or spurious peaks not related to signal components.

\subsection{First guess estimation of microphone and sources positions}
Since the direct path is the shortest among all the possible paths of the signal in the room, the first TOA will correspond to the direct path for every couple of microphone and source. Moreover, as the direct path is likely the strongest one, it will be almost surely detected by the peak finder. Thus, in order to find the TOAs related to direct path, it is sufficient to select the delays  corresponding to the first detected peak $\tilde{\tau}^{a,n}_{m,1}, \hspace{0.2cm} n=1,\hdots,N; \hspace{0.2cm} m = 1,\hdots M$.

From the direct path TOAs it is possible to infer the position of real sources and microphones also with unkonwn emission and offset times. To do this we rely on the method developed in \cite{Gaubitch2013}, inspired by the rank based method of \cite{Crocco:etal:2012}, here below briefly described.

As a first step let us convert the TOAs into distances 
\begin{equation}
d^{a,n}_{m}=c\tilde{\tau}^{a,n}_{m,1}
\end{equation}
Then, arrange the squared distances into a $N\times M $ matrix $\mat D$ such that:
\begin{equation}
\mat D =(d^{a,n}_{m})^2-(d^{a,1}_{m})^2-(d^{a,n}_{1})^2
\end{equation} 
It can be easily demonstrated \cite{Crocco:etal:2012} that, if $\tau^n_e=0$ and $\tau^o_m=0$ for all $n$ and $m$ and no  error is present on the estimated TOAs the matrix $\mat D$ has rank equal to 3 and can be written as: 
\begin{equation}
\mat D = \mat B^{\top}\mat S
\end{equation}
where $\mat B$ and $\mat S$ are $3 \times N$ and $3 \times M$ matrices contain the 3D coordinates of the sources and microphones respectively. Thus, the matrices $\mat S$ and $\mat B$ can be recovered from the left and right singular vectors of the Singular Value Decomposition of $\mat D$  truncated to the third singular value. Since the relation $\mat D = \mat B^{\top}\mat Q\mat Q^{-1}\mat S$ holds for any invertible $3 \times 3$ matrix $\mat Q$, nine additional parameters have to be estimated, to recover the source and microphone coordinates. To this end a suitable cost function exploiting the quadratic relations obtained squaring the relations in (4)
(\ref{dist}) is described in \cite{Crocco:etal:2012}. 
In the general case in which emission and offset times are different from zero the rank 3 relationship holds for a different matrix $\hat{\mat D}$, given by the sum of $\mat D$ and a correction matrix $\mat{\phi}$ function of $\tau^n_e$ and $\tau^o_m$. This results in the following iterative algorithm \cite{Gaubitch2013}:  
\begin{center}
\begin{algorithm}
	\caption{Iterative algorithm to estimate sources and microphones positions, emission times and offset times}
	\label{alg:Gaubitch}
	\begin{algorithmic}[1]
		\STATE initialize $\mat{\phi}(\tau^n_e,\tau^o_m)$ and compute $\hat{\mat D}=\mat D+\mat{\phi}(\tau^n_e,\tau^o_m)$
		\REPEAT
		\STATE find $\hat{\mat D}_3$ as the closest rank 3 approximation of $\hat {\mat D}$
		\STATE find the residual $\mat E= \hat{\mat D}_3-\hat{\mat D}$
		\STATE estimate $\tau^n_e,\tau^o_m$ from $\mat E$ by linear least squares.
		\STATE update matrix $\mat{\phi}(\tau^n_e,\tau^o_m)$
		\STATE compute the new matrix  $\hat{\mat D}=\mat D+\mat{\phi}(\tau^n_e,\tau^o_m)$
		\UNTIL {$\|\mat E\|_F$ lower than a threshold}
		\STATE  Estimate $\vec b^n$ and $\vec s_m$ given  $\hat{\mat D}_3 =  \mat B^{\top}\mat Q\mat Q^{-1}\mat S$, solving for the matrix $\mat Q$
	\end{algorithmic}
\end{algorithm}
\end{center}
The above algorithm outputs the set of estimated microphone positions $\tilde{\vec s}_m$, source positions $\tilde{\vec b}^n$, emission times $\tilde{\tau}^n_e$ and offset times $\tilde{\tau}^o_m$. 
The estimated TOFs can be obtained by a simple subtration:
\begin{equation}
\tilde{\tau}^n_{ml} = \tilde{\tau}^{a,n}_{ml}-\tilde{\tau}^n_e-\tilde{\tau}^o_m.
\end{equation}
Notice that in the following we denote with a tilde the estimation of a given parameter.

\subsection{Walls estimation by greedy iterative approach}
From the previous section we have obtained, for each microphone $n$ and each source $m$ a set of estimated TOFs $\mathcal{T}^n_{m0}=\left\{\tilde{\tau}^n_{m1},\tilde{\tau}^n_{m2},\ldots,\tilde{\tau}^n_{mL(n,m)}\right\}$ of length $L(n,m)$. Moreover, we have estimated as well the set of sources and microphones positions $\tilde{\vec b}^n$, $\tilde{\vec s}_m$.

A straightforward method to find the correct reflector position is to build a cost function given by the sum of the absolute differences of each TOF, function of the position reflector through equations (\ref{i1}), (\ref{i2}), (\ref{eq:image}), (\ref{dist}) and each each TOF estimated from the signal. In the following we will call these two TOFs the geometry TOF and the signal TOF respectively. The underlying idea is that when all the reflectors are in the right position the two set TOFs perfectly match, leading to a global minimum in such cost function. 

Unfortunately, there are three important issues to deal with: the first is the nonlinear nature of (\ref{i2}) and (\ref{eq:image}) that may give rise to a nonconvex cost function and consequently local minima. The second is the presence of spurious peaks or missing peaks in the estimated TOF from signal. Finally, the third is the well known labelling problem or echo sorting problem, i.e. the unknown association between TOFs from signals and related reflectors. Thus, we need a method to find a first guess solution to initialize the cost function, a method to ignore spurious peaks and make the cost function robust to missing data and a labelling procedure.  

A brute force approach to solve the labeling problem would be to consider all the possible matchings between TOFs from geometry and TOFs from signal and minimizing, for each set of matchings, a cost function function of the reflectors positions. This approach is computationally infeasible due to the combinatorial explosion of possible matchings. 

An alternative approach to bypass the above three issues would be to discretize the space of all possible reflector positions and perform a brute force search over all the possible positions combinations, checking, for all combinations, if the TOFs given by the geometry correctly match the estimated TOFs from signals. In this case the problems of labeling, missing data and spurious peaks is solved simply by considering, for each TOF from geometry, the nearest TOF from signal, and applying opportune robust cost functions to limit the effect of missing and spurious peaks. 

Moreover, notice that the problem of local minima is implicitly solved by exhaustive space over all the possible solutions. Unfortunately, also this procedure is computationally infeasible due to dimensionality of the search space equal to $3K$. However, if we initially consider only the first order reflections, we can decouple the search for each single reflector, decreasing in this way the dimensionality of the search space to $3$.   

Then, once the first reflector position is estimated, the image sources associated to it can be exploited to infer the subsequent reflectors positions. This results in a greedy iterative algorithm, where at each iteration a new reflector position is estimated and second order reflections are incrementally added to the data.

In more details, let us discretize the 3D euclidean space of possible locations of the planar surfaces in a 3D cartesian grid $\vec r^{grid}_i = (x_i,y_i,z_i)$ with $i=1,\ldots,I$.
The grid boundaries are set according a coarse guess of the dimension of the room and the grid spacing is given by the required precision and the computational resources available. Now, 
rename for convenience the estimated real source positions $\tilde{\vec b}^n$ with  $\tilde{\vec p}^n_0$.  Moreover, consider the index $j=1,\ldots,K$ as the iteration index of the algorithm and $k = 0,\ldots,j-1$ as the index of image sources (and real ones for $k=0$) that will be be progressively added along with the iterations.

Consider the first iteration for which $j=1$ and $k=0$.
For a given real source $n$, a microphone $m$ and a tentative reflector position $r^{grid}_i$, the  time of flight $\tau^{geom,n}_{mk}(i)$ computed from the geometry of the problem is given by:  
\begin{equation}
	\tau^{geom,n}_{mk}(i)=\|\tilde{\vec s}_m- IM(\tilde{\vec p}^n_k,\vec r^{grid}_i)\|_2/c.
	\label{g1}
\end{equation}
For such TOF, we search for the index $\tilde{l}$ of the closest TOF estimated form the signal, with a nearest neighbours (NN) approach:
\begin{equation}
	\tilde{l}(n,m,k,i) = NN(\mathcal{T}^n_{mj},\tau^{geom,n}_{mk}(i)).
	\label{g2}
\end{equation}
Now, we need a score function to evaluate the goodness of the matching between the two delays $\tau^{geom,n}_{mk}$ and $\tilde{\tau}^n_{m,\tilde {l}(n,m,k,i)}$. Such score should be rapidly decaying whenever the two delays are too distant, since in this case the guessed reflector position is probably wrong. At the same time the score should be robust to missing TOFs: if the peak finder fails to detect a peak from the signal, the nearest neighbour procedure identifies a matching within arbitrarily distant TOFs. A suitable score function $S^n_{mk}(i)$ is the following:
\begin{equation}
	S^n_{mk}(i)=exp\left(-\frac{\left(\tau^{geom,n}_{mk} -\tilde{\tau}^n_{m,\tilde {l}(n,m,k,i)}\right)^2}{2\sigma^2}\right)+\epsilon.
	\label{g3}
\end{equation}
Two parameters are present: the standard deviation $\sigma$ of the Gaussian and the threshold parameter $\epsilon$. The first one rules the rate of decay of the score function and should be set according to the expected standard deviation of the peak finder error (i.e. its precision). The second one imposes a lower bound on the score function, so making it robust to missing TOFs.
The total score function $S^{tot}(i)$ is given by the product of all the score functions related to the couples of microphones and real sources as follows:
\begin{equation}
	S^{tot}(i)= \prod_{n=1}^N \prod_{m=1}^M\prod_{k=0}^{j-1}S^n_{mk}(i).
	\label{g4}
\end{equation}
Recall that at first iteration $j=1$ and $k=0$, so the last product structure is not relevant.
Finally we search by a brute force approach for the reflector position yielding the maximum score: 
\begin{equation}
	\tilde{i} = \arg \min_i S^{tot}(i),
	\label{g5}
\end{equation}
\begin{equation}
	\tilde{\vec r}_{j}=\vec r^{grid}_{\tilde{i}}
	\label{g6}
\end{equation}
with $j = 1$.
In this way we have found the first estimated reflector position $\tilde{\vec r}_1$. 
In order to perform the search of the second reflector, starting iteration $j=2$, we need to remove from the set of estimated TOFs all the TOFs that are likely related to the first reflector, otherwise the search procedure would fall again in the same global maximum already found for the first reflector. To do this we subtract to each set of estimated TOFs $\mathcal{T}^n_{m,j}$ , the set of TOFs already matched with the geometry TOFs calculated from the first reflector $ \mathcal{T}^{matched,n}_m$, as follows: 
\begin{equation}
	\mathcal T^n_{m,j+1} = \mathcal{T}^n_{m,j} \setminus  \mathcal{T}^{matched,n}_m,
	\label{g7}
\end{equation}
where $\mathcal{T}^{matched,n}_m$ is defined as the set of TOFs for which at least a geometry TOF is distant less than a predefined threshold $thr_3$:
\begin{equation}
	\mathcal{T}^{matched,n}_m = \left\{ \tilde{\tau}^n_{ml} \mid \left( \exists k \mid | \tau^{geom,n}_{m,k}(\tilde{i})-\tilde{\tau}^n_{ml}|< thr_3 \right)   \right\}
	\label{g8}
\end{equation}
The threshold $thr_3$ is necessary to avoid the undesired removal of TOFs $\tilde{\tau}^n_{ml}$ too distant from the corresponding geometry TOFs. In this case in fact the matched signal TOFs are unlikely to belong to the current reflector, but they are likely due to missing data.
Finally, once the first reflector $\tilde{\vec r}_1$ has been estimated, we can consider the corresponding $n$ image sources $\tilde{\vec p}^n_1$ for the search of the second reflector. In practice this means to take into account the second order reflections between the already estimated reflector an the next one to be estimated. 

Thus for each of the pair $n,m$ of matches, there will be searched one for the real source $\tilde{\vec p}^n_0$ and one for the virtual source  $\tilde{\vec p}^n_1$. Notice that all the previous formulas are still valid by simply increasing the iteration index $j$ from $1$ to $2$. By repeating the procedure, the second reflector is estimated and the new image sources are added, so making the estimation more and more robust. Notice that the first estimated reflector, chosen as the one yielding the maximum score at iteration $1$, will be the one for which the peaks are most accurately estimated and few missing data are present. Thus second order reflections, that are not considered at all in the algorithm first iteration, are likely not necessary to obtain an accurate estimation of the first reflector position. 
Going on with the iterations, the reflectors will become more and more difficult to estimate due to the less reliable peaks and increased missing data However, this effect is compensated by the increasing amount of data provided by second order reflections given by the first estimated reflectors. In a sentence, the reflectors providing the strongest reflected signals 
help with their image sources the estimation of the other ones characterized by weaker reflected signals. 

At the generic iteration $j$, $j-1$ image sources and a real source are considered to estimate the delays $\tau^{geom,n}_{m,k}(i)$ using Eq. (\ref{g1}). Subsequently $NMJ$ matchings are evaluated for each position $i$ of the grid in (\ref{g2}) and $NMJ$ score functions are computed by (\ref{g3}) and combined in (\ref{g4}). Once the $j-th$ reflector has been estimated through (\ref{g5}) and (ref{g6}) a variable number of delays $<j$ is removed from each of the $NM$ delay sets $\mathcal{T}^n_{m,j+1}$, as described in (\ref{g7}), (\ref{g8}). Finally a new set of $N$ image sources $\tilde{\vec p}^n_{j}$ is considered for the computation of geometry delays in (\ref{g1}) at iteration $j+1$. 

\subsection{Global refinement}

Given this initial estimation of the $K$ reflectors $\tilde{\vec r}_k$, we can now proceed to globally refine the solution. 
A preliminary step is to assure that the correct TOFs labelling is achieved by pruning out all spurious peaks and taking into account missing data. To do this, let us compute all the geometry TOFs related to both direct path, first and second order reflections for all the estimated reflectors, real sources and microphones. With slight abuse of notation, we define:
\begin{equation}
	\tau^{geom}(\vec b^n,\vec s_m,\vec r_k,\vec r_z) = \|{\vec s}_m- {\vec b}^n\|_2/c,
\end{equation}
with $n = 1,\hdots,N $, $m=1,\hdots,M$, $k=0$, $z = 0$  for direct path TOFs;
\begin{equation}
	\tau^{geom}(\vec b^n,\vec s_m,\vec r_k,\vec r_z) = \|{\vec s}_m- IM(\vec b^n,\vec r_k)\|_2/c,
\end{equation}
with  $n = 1,\hdots,N $, $m=1,\hdots,M$, $k=1,\hdots,K$, $z = 0$ for first order TOF;
\begin{equation}
	\tau^{geom}(\vec b^n,\vec s_m,\vec r_k,\vec r_z) = \|{\vec s}_m- IM(IM(\vec b^n,\vec r_z),{\vec r}_j)\|_2/c,
\end{equation}
with $n = 1,\hdots,N $, $m=1,\hdots,M$, $k=1,\hdots,K$, $z =1,\hdots,k-1,k+1,\hdots,K$ for second order TOFs.

Then, we search for the Nearest Neighbours (NN) among the sets of of signal TOFs $\mathcal{T}^n_m$
\begin{equation}
	\tilde{l}(n,m,k,z) = NN(\mathcal{T}^n_m,\tau^{geom}(\vec b^n,\vec s_m,\vec r_k,\vec r_z))   
\end{equation}
To prevent the negative effect of missing data we check when the NN distance exceeds a threshold $thr_3$ \footnote{Notice that $thr_3$ is not a new threshold : It is introduced in Eq. \ref{g8} with the same meaning} and set a binary indicator function:
\begin{eqnarray} 
	Ind_{n,m,k,z} =  \nonumber \\
=	\begin{cases}
		1, & \text{if}\ | \tau^{geom}(\vec b^n,\vec s_m,\vec r_k,\vec r_z) - \tilde{\tau}^n_{m,\tilde{l}(n,m,k,z)}|< thr_3    \\
		0, & \text{otherwise}  
	\end{cases}
\label{sec4_1}
\end{eqnarray}
Moreover, for the sake of compactness of notation, we define $Ind_{n,m,k,z}=0$ for the indices combinations that do not correspond to any delays i.e. $k = 0 \wedge z \neq 0$  and $k=z \wedge z>0 \wedge k>0$.
We recover the pruned set of TOAs by summing to the pruned TOFs the estimated emission and offset times:
\begin{equation}
	\tilde{\tau}^{a,n}_{m,\tilde{l(n,m,k,z)}}=\tilde{\tau}^{n}_{m,\tilde{l(n,m,k,z)}}+\tilde{\tau}^n_e+\tilde{\tau}^o_m
\end{equation}

Now we can solve the global problem, instantiating the non-linear Least Squares cost function:
%
%
\begin{eqnarray}
	\min_{\vec b^n, \vec s_m, \vec r_k \tau^e_n, \tau^o_m } \sum_{n=1}^N\sum_{m=1}^M\sum_{k=0}^K\sum_{z=0}^K Ind_{n,m,k,z}\cdot \nonumber \\ \cdot\left( \tilde{\tau}^{a,n}_{m,\tilde{l}(n,m,k,z)}-\tau^{geom}(\vec b^n,\vec s_m,\vec r_k,\vec r_z)-\tau^n_e-\tau^o_m  \right)^2  \nonumber \\ 
\end{eqnarray}
with a Matlab standard solver (e.g. lsqnonlin) based on the Levemberg-Marquardt method.
The variables optimized are initialized with $\tilde{\vec b}^n,\tilde{\vec s}_m,\tilde{\vec r}_k,\tilde{\tau}^n_e,\tilde{\tau}^o_m$ as estimated  
in the previous sections.

\section{Synthetic tests}
A set of synthetic experiments was performed in order to assess the accuracy and robustness of the method in different operative conditions including: Variable number of microphones, room shape and signal to noise ratio on the received audio signals. 
Concerning the room shape, we generated convex polyhedra with six faces and arbitrary shape, in order to depart from the simpler shoebox model of a room. In particular, we started from a cubic room of side equal to $6m$ and randomly perturbed the length and orientation of the normal to each plane, according to two random variables of uniform distribution between $-1.5m$  and $1.5 m$ for the normal length and $-20$ and $20$ degrees for the normal orientation. Microphones and sources ground truth positions were generated according to a uniform distribution inside the room volume. We run the algorithm over $30$ different rooms and for each room we set a number of microphones equal to $12$ and $8$ with a number of sources equal to $20$.   

\begin{figure}[bth]
	\begin{center}
		\hspace{-0.3cm}
		\includegraphics[width=0.53\linewidth]{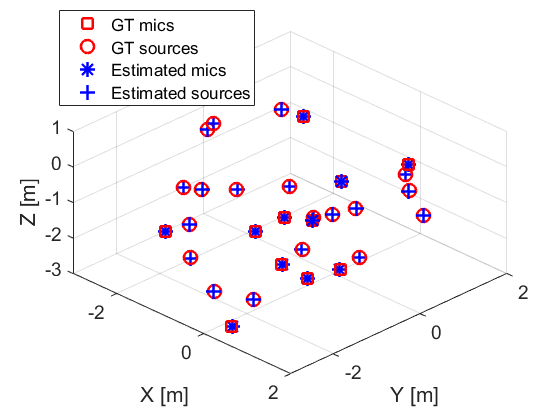}
		\hspace{-0.6cm}
		\includegraphics[width=0.53\linewidth]{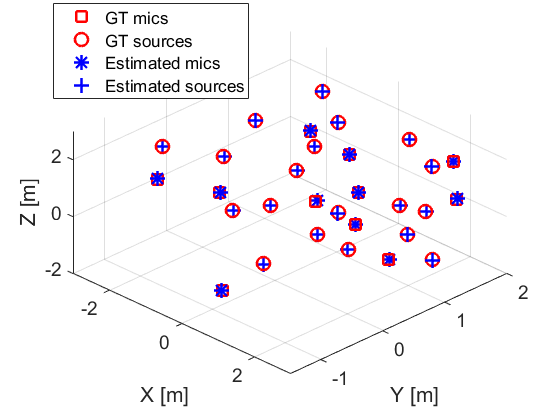} \\	
	\end{center}
	\vspace{-.35cm}
	\caption{Two examples of microphone and source positions estimation and respective ground truth for SNR = 14dB (a) and SNR = -6 dB (b). Legend: blue crosses (estimated sources); red circles (ground truth sources); blue stars (estimated microphones); red squares (ground truth microphones).}
	\label{fig:synthms}
	\vspace{-.2cm}
\end{figure}

\begin{figure}[bth]
	\begin{center}
		\hspace{-0.3cm}
		\includegraphics[width=0.53\linewidth]{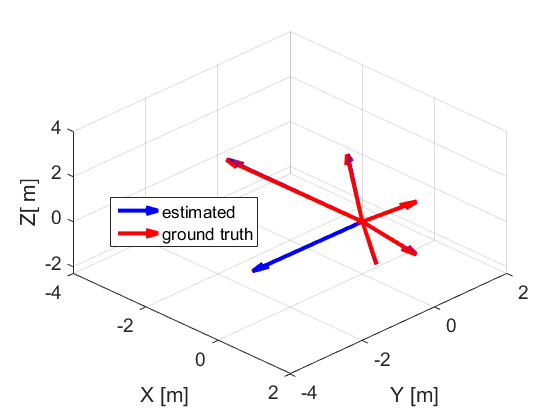} \hspace{-0.6cm}	\includegraphics[width=0.53\linewidth]{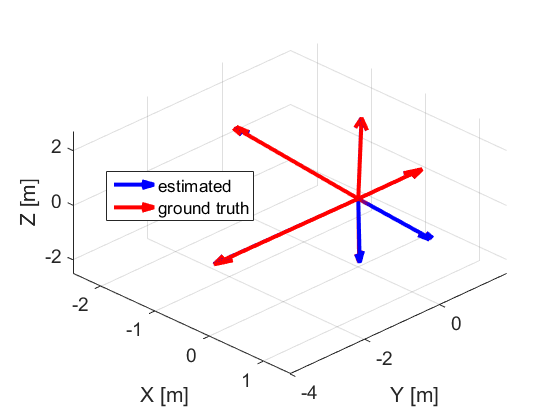}  \\
		(a)	\hspace{3.5cm} (b)\\
		
	\end{center}
	\vspace{-.35cm}
	\caption{Two examples of room reconstruction for SNR = 14 dB (a) and SNR = -6 (b): each vector represents the normal to a plane whose distance from the coordinate center is given by the vector length. Legend: blue (estimated plane); red (ground truth plane).}
	\label{fig:synth}
	\vspace{-.2cm}
\end{figure}

\begin{figure*}[thb]
	\begin{center}
		\includegraphics[width=0.26\linewidth]{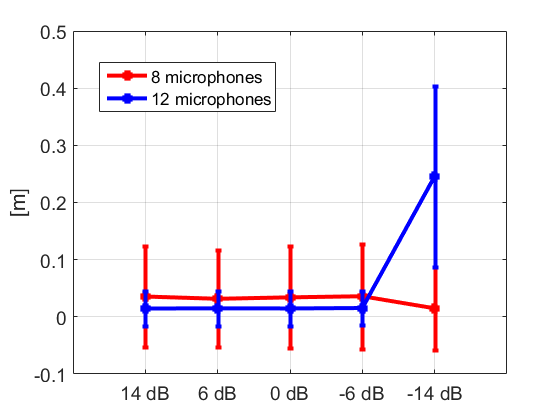} \hspace{-0.6cm}
		\includegraphics[width=0.26\linewidth]{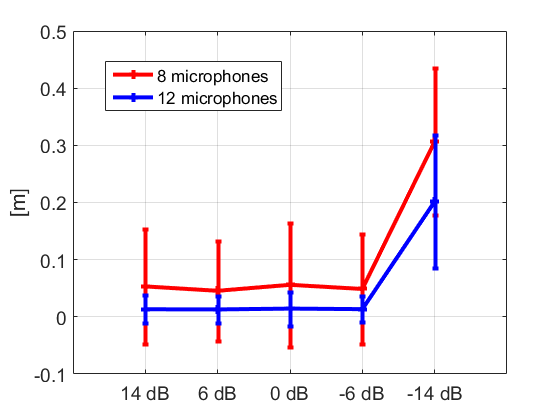} 
		\hspace{-0.6cm} 
		\includegraphics[width=0.26\linewidth]{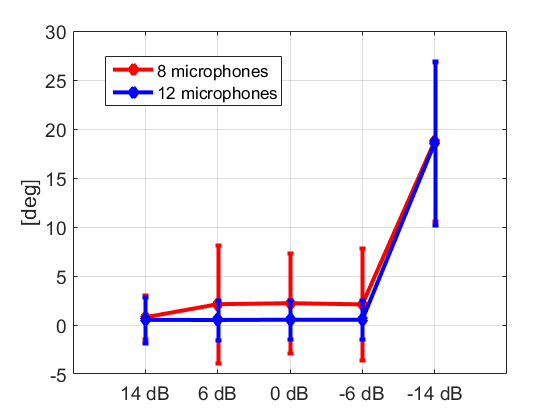} 
		\hspace{-0.6cm}
		\includegraphics[width=0.26\linewidth]{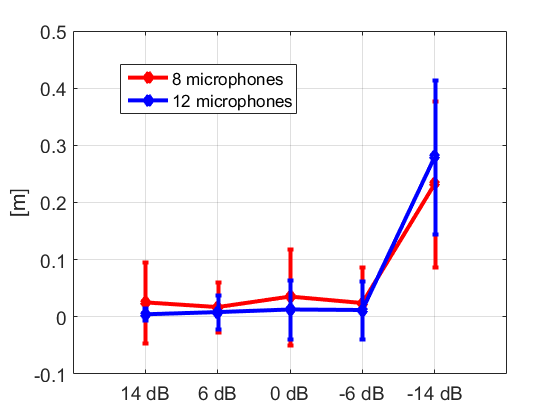} \\ 
		(a) \hspace{3.8cm} (b) \hspace{3.8cm} (c) \hspace{3.8cm} (d)
	\end{center}
	\vspace{-.35cm}
	\caption{Mean and standard deviations of reconstruction errors versus SNR (inliers) for 8 (red) and 12 microphones (blue). Figure (a) shows the microphone position errors while Figure (b) presents the source position errors. Regarding the room reconstruction performance, Figure (c) shows the angle error over the normals to the surfaces and Figure (d) presents the distance error on normal to the planes.}
	\label{fig:stat1}
	\vspace{-.2cm}
\end{figure*}

\begin{figure*}[thbp]
	\begin{center}
		\includegraphics[width=0.26\linewidth]{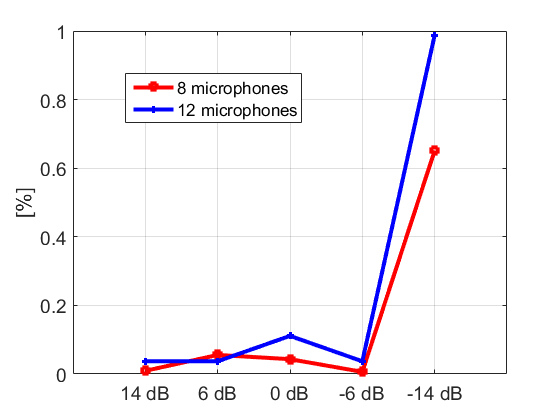}
		\hspace{-0.6cm}
		\includegraphics[width=0.26\linewidth]{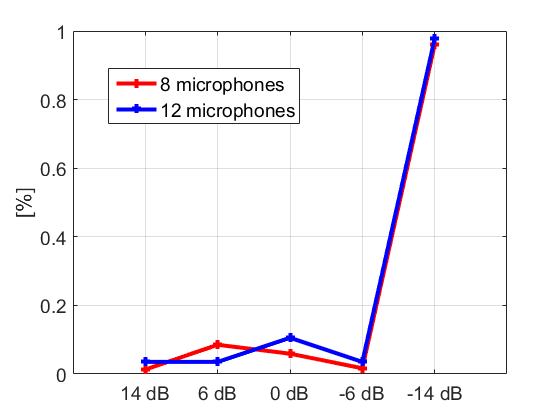}  
		\hspace{-0.6cm}
		\includegraphics[width=0.26\linewidth]{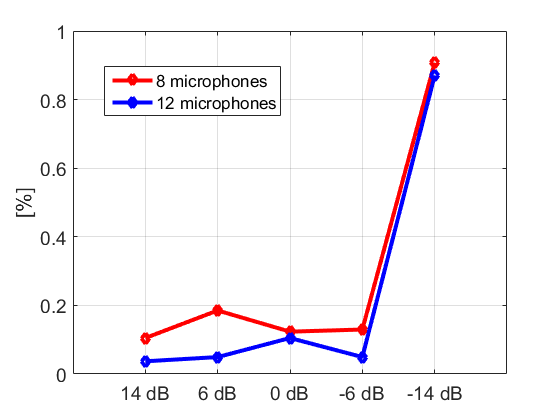} 
		\hspace{-0.6cm}
		\includegraphics[width=0.26\linewidth]{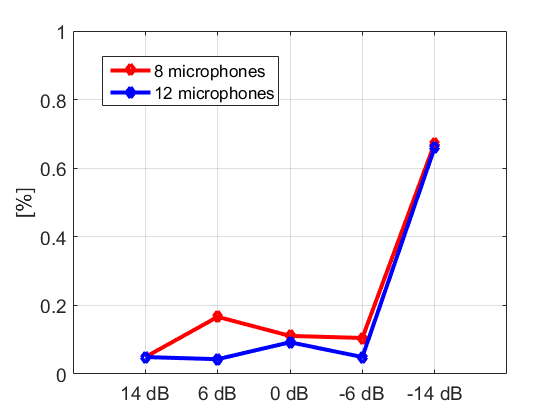} \\ 
		(a) \hspace{3.8cm} (b) \hspace{3.8cm} (c) \hspace{3.8cm} (d)	
	\end{center}
	\vspace{-.35cm}
	\caption{Percetnage of outliers  reconstruction errors versus SNR for 8 (red) and 12 microphones (blue). Figure (a) shows the microphone position errors while Figure (b) presents the source position errors. Regarding the room reconstruction performance, Figure (c) shows the angle error over the normals to the surfaces and Figure (d) presents the distance error on normal to the planes.}
	\label{fig:stat2}
	\vspace{-.2cm}
\end{figure*}
As transmitted signals we considered a chirp of length $5s$ and frequency sweep from 3 kHz to 6 kHz. Notice that the relative bandwidth of the chirp was not so high in order to comply with the limited bandwidth of the speaker used in real experiments. In fact, we intentionally employed low budget speakers and microphones in order to show that the reconstruction method does not require high-end equipment.  
The room impulse responses were computed adopting the image method of Allen and Berkley \cite{Allen1979}. A white noise was added to each received signal in order to reproduce different levels of signal to noise ratio, namely the following SNR were adopted: $14dB$, $6dB$, $0dB$, $-6dB$, $-14dB$. This stressed the algorithm to work with very high noise conditions.

Concerning the algorithm parameters, the thresholds related to peak finding  $thr_1$ and $thr_2$ in Algorithm 1 were set to $0.9$ and $0.1$ respectively for all the tests. The step of the grid used to discretise the space of the possible locations of surfaces was set to $0.2$ m, the threshold $\epsilon$ and the standard deviation $\sigma$ in Eq. (\ref{g3}) were set to $0.1s$ and $1.47 \cdot 10^{-4}s$ (corresponding to $0.05m$) respectively.
Finally threshold $thr_3$ in Eq. (\ref{g8}) and  Eq. (\ref{sec4_1}) was set to $5.88 \cdot 10^{-4}s$ (corresponding to $0.2m$). 

The average computational load for a whole reconstruction on a 2.6 GHz CPU and with a Matlab implementation is about $258s$ for $20$ sources and $12$ microphones. In details, the compression of signals with the matched filter took about $6s$, the peak finding procedure $2s$, the computation of the initial solution for microphone and sources about $40s$, the first guess estimation of surfaces by grid search about $150s$ and the final global refinement about $60s$. Notice that the grid search algorithm, despite its ``brute force'' nature, has a computational time of the same order of magnitude given by the global refinement and the initial estimation of microphones and sources positions. This fact demonstrates that the proposed algorithm allows to tackle the fully uncalibrated room reconstruction problem without any heuristics aimed at simplifying the peak labeling problem and avoiding at the same time the combinatorial explosion that is most evident in previous approaches.

Regarding microphones and sources, accuracy of reconstruction was evaluated by computing the Euclidean distance between ground truth and estimated 3D positions, after Procrustes alignment. This was necessary because the 3D reconstruction is up to an unknown rigid transformation, if no other a priori information about the geometry of the sensors/sources is given. Concerning room surfaces, we evaluates the vectors normal of each plane whose length is given by the distance from the plane to the coordinate center that coincides with the room center. Given such normal vectors, we compute the angle between ground truth and estimated vectors and their absolute difference in length . 

In Figs. \ref{fig:synthms} and \ref{fig:synth} two examples of room reconstructions and the respective  microphone and source estimation are displayed, for  two different SNR = 14 dB and SNR =  6 dB. As it can be seen, a nearly exact reconstruction is obtained with both the SNR values.
The average error versus SNR, for microphones sources and surfaces is displayed in Figs. \ref{fig:stat1}
and \ref{fig:stat2}. In particular, to decouple the effect of outliers from the accuracy of inliers we displayed in Fig.\ref{fig:stat1} the mean and standard deviation of the four errors (microphone and source displacement, angle and distance from center of surfaces) limiting the computation to inliers and we displayed in Fig. \ref{fig:stat2} the percentage of outliers. We considered as inliers estimated microphones, sources and surface distances from center whose distance from ground truth is below $0.5 m$ and surface angles whose absolute difference with ground truth is below $20$ degrees.  All data are displayed versus the SNR and for a number of microphones equal to $8$ and $12$. Concerning inliers, it can be noted that error is nearly constant for an SNR ranging from $14 dB$ to $-6 dB$ while abruptly increases for SNR = $-14 dB$. Also the percentage of outliers follows the same behaviour.  
This is due to the fact that, above an SNR threshold placed between $-6dB$ and $-14 dB$, noise  is largely pruned out by the peak finder and it minimally affects the peaks positions. Moreover the inherent robustness of the algorithm is able to cope with mis-detected peaks (these peaks are also present in condition of zero noise due to side lobe peaks of the compressed signal and to very close times of arrival whose peaks tend to merge together). Below this SNR  threshold, the noise amplitude becomes comparable to the amplitude of signal peaks. In such case, the peaks found by the detector will belong in great part to noise so making data practically useless. 

The comparison between $8$ and $12$ microphones shows that, as expected, results are generally superior and more stable with $12$ microphones. Look in particular at the standard deviation of inliers that are considerably larger for $8$ microphones. 

Overall, when the SNR is kept at a reasonable level, a very good average accuracy is achieved. In particular the average accuracy is about: $0.015m$ ($12$ microphones) and $0.035m$ ($8$ microphones) for microphones; $0.015m$ ($12$ microphones), and $0.05m$ ($8$ microphones) for sources,  $0.5 deg$ ($12$ microphones), and $2deg$ ($8$ microphones) for surface angles, $0.01m$ ($12$ microphones), and $0.02m$ ($8$ microphones) for surface distances from the center.  

The number of outlier is kept below $10\%$ for all the errors for $12$ microphones and below $20\%$ for $8$ microphones. It is interesting to note that the great part of outliers is related to a few trials in which the starting guess reconstruction of microphones and sources is grossly inaccurate, so affecting the subsequent estimation of planar surfaces. In such cases the magnitude of the starting error does not allow the final refinement to effectively recover the right solution. 
%





\section{Real experiments}
\subsection{Setup}

Evaluating room reconstruction algorithms in a real environment is a very complex task. This is mainly due to the difficulties and equipment costs in obtaining reliable ground truth data. To help the advancements in this field, here we make public a dataset that includes precise positions of microphones and sources, 3D scans of a room and the signal emitted by sources\footnote{Dataset with ground truth can be freely downloaded at: www.iit.it/datasets/vgm-3d-room-reconstruction-dataset}. This allows to perform accurate evaluation of the proposed algorithm performance.

\begin{figure}[bth]
	\begin{center}
		\includegraphics[width=.7\linewidth]{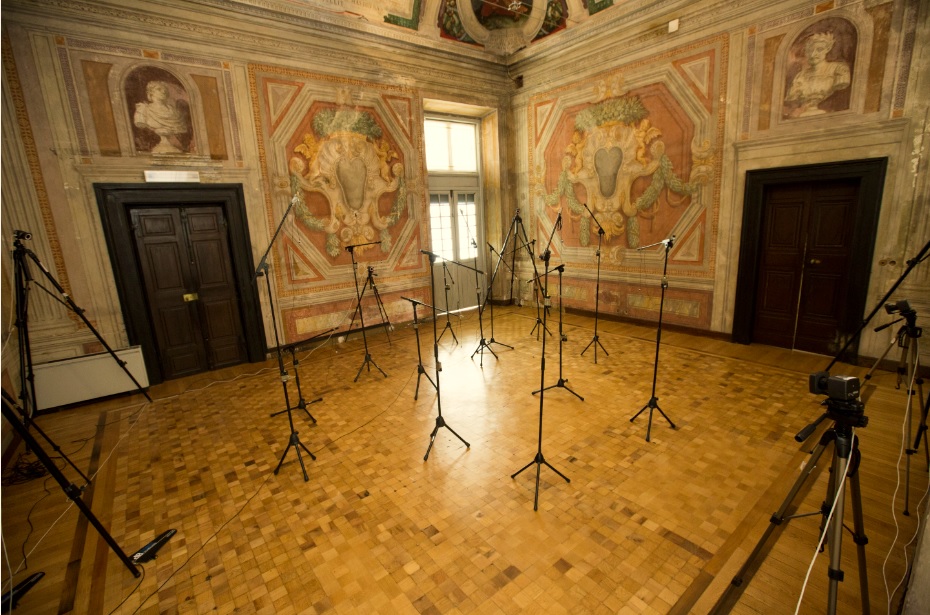}   
	\end{center}
	\vspace{-.35cm}
	\caption{View of the room selected for experimental tests with the deployed microphones, source and motion capture system.}
	\label{fig:room}
	\vspace{-.2cm}
\end{figure}

\begin{figure}[bth]
	\begin{center}
		\includegraphics[width=.7\linewidth]{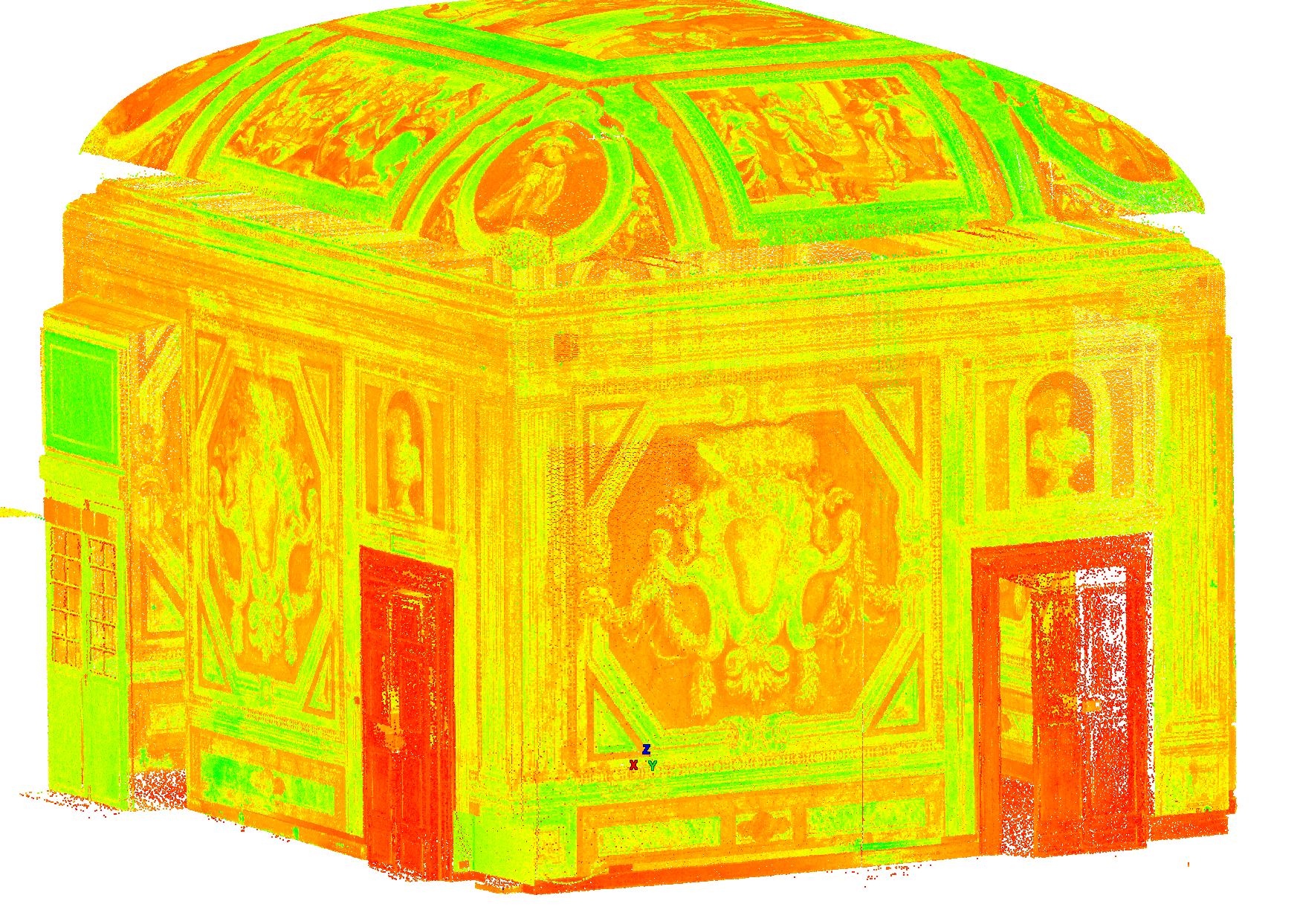}  
	\end{center}
	\vspace{-.35cm}
	\caption{High resolution 3D point cloud reconstruction of the room using a LEICA C-10 laser scanner.}
	\label{fig:laser_scanner}
	\vspace{-.2cm}
\end{figure}

The location selected for experimental tests was a room in a sixteenth century mansion part of UNESCO heritage in Genoa. The room, displayed in Fig. \ref{fig:room}, was rectangular with a vaulted ceiling and wooden floor, with dimensions of about $8.5m \times 7.5m \times 7m$.
We displaced $12$ radio frequency omnidirectional Lavalier microphones in the central part of the room, inside an area of about $3m \times 3m$ . Each microphone hanged from a tripod with different heights (from $20cm$ to $2m$). As audio sources we used a small wireless speaker of about $3$ cm diameter (VEHO360). The speaker was moved in 17 different locations in the room and the transmitted signal was the same chirp already used for the synthetic experiments. The signals from the microphones were synchronously acquired and digitally converted by a National Instrument PXI system and finally stored in a PC. Notice, as an additional difficulty, a significant amount of noise was present due to the traffic in a street nearby.

In order to properly match the acquired signals, the matched filter is obtained by transmitting once more the same signal and placing  a microphone very close to the source in an anechoic environment: in this way both the acquired signals and the matched filter are convoluted by the same microphone and loudspeaker impulse responses. 
\begin{figure}[tbhp]
	\begin{center}
		\includegraphics[width=.45\linewidth]{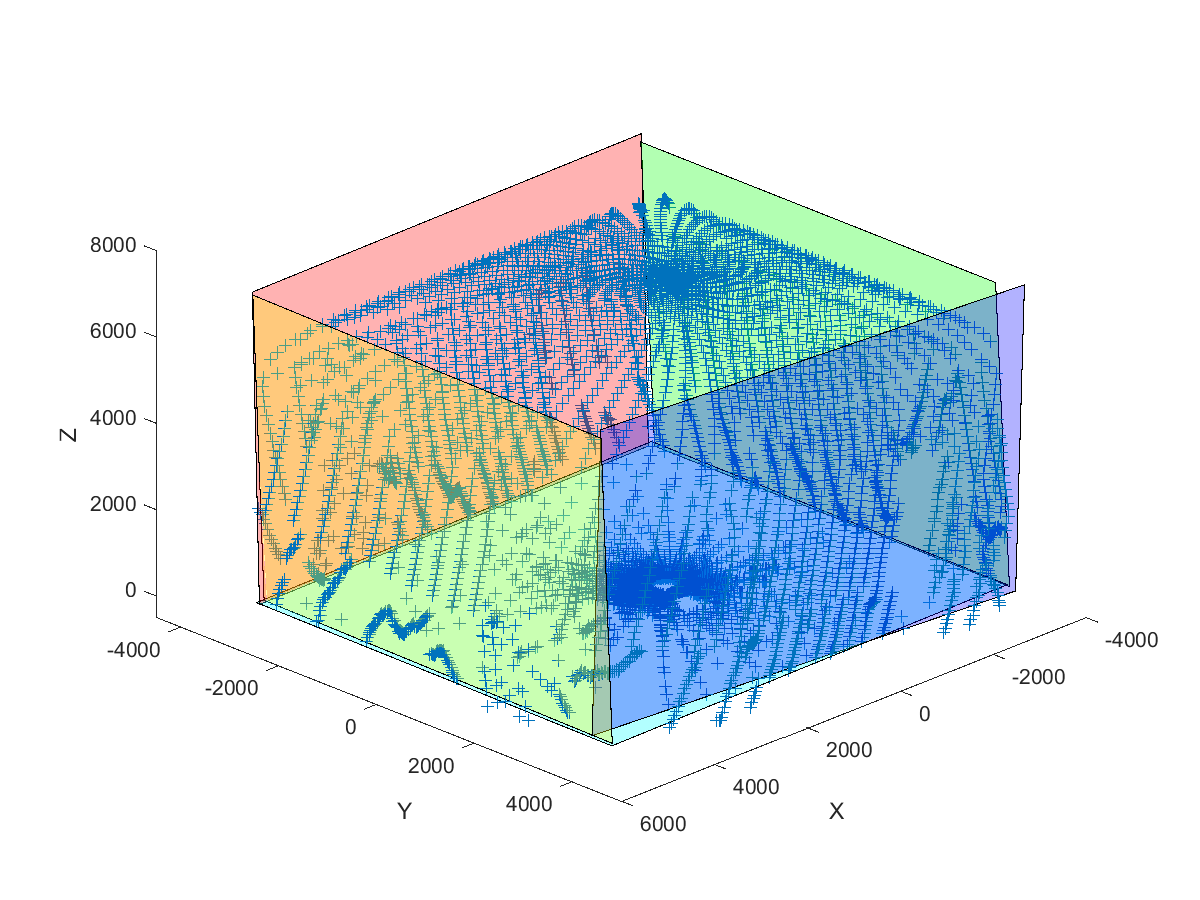}  
		\includegraphics[width=.45\linewidth]{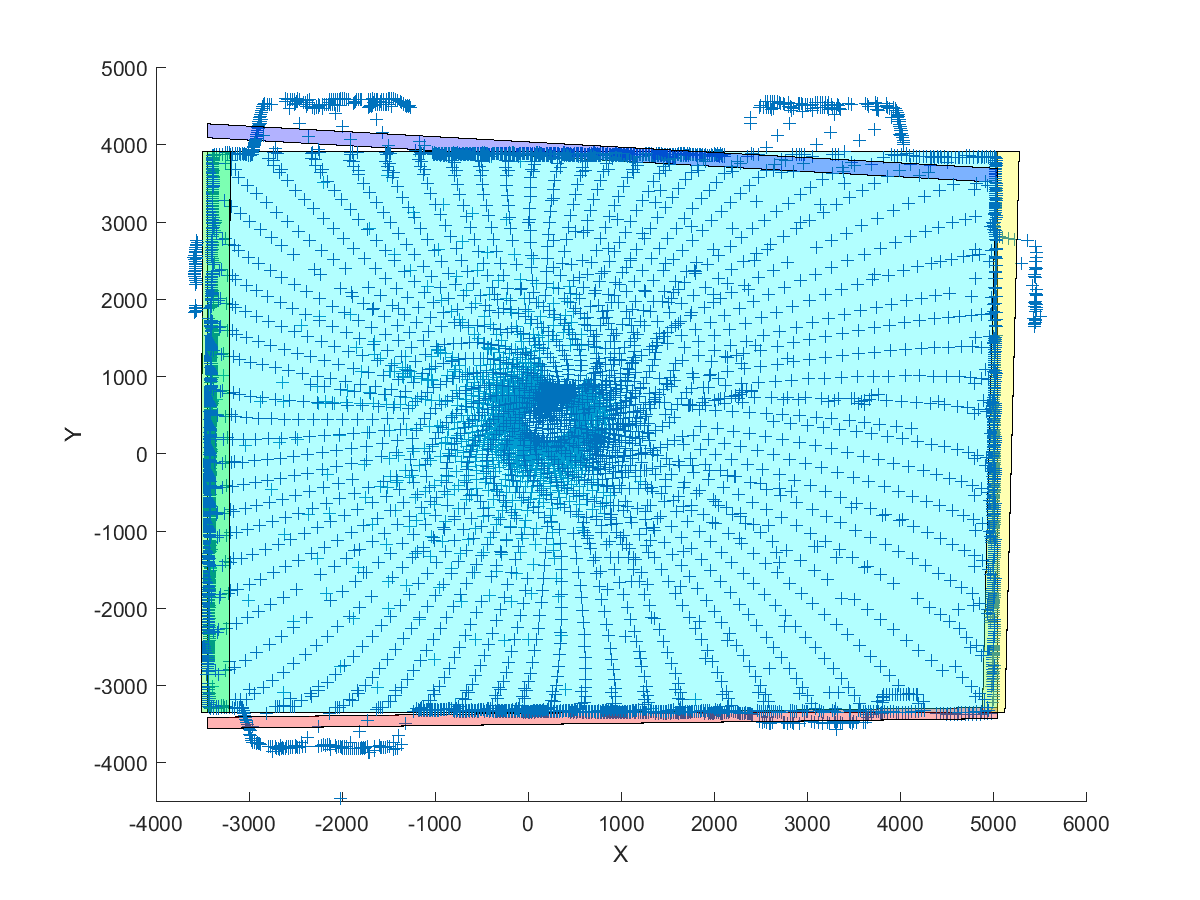} \\
		(a) \hspace{3.5cm} (b) \\
		\includegraphics[width=.45\linewidth]{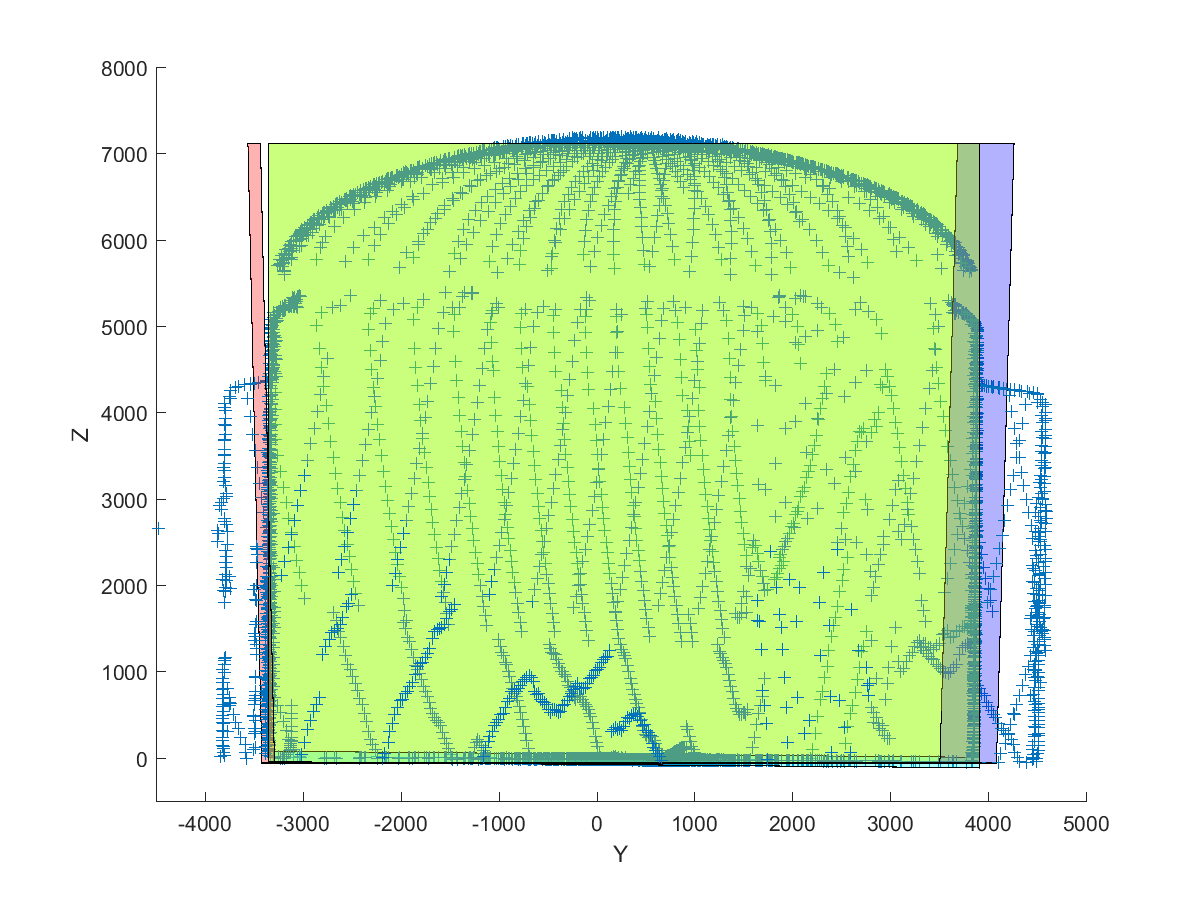} 
		\includegraphics[width=.45\linewidth]{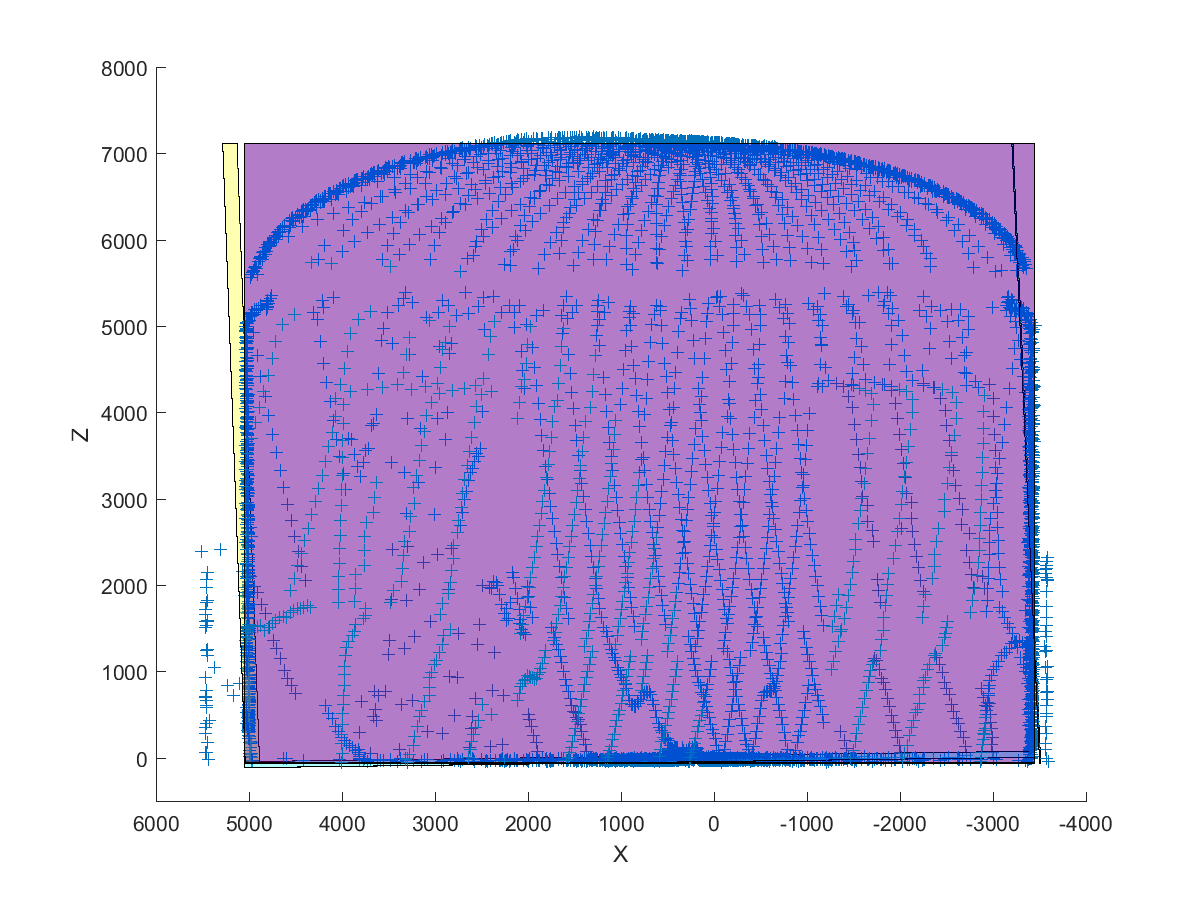}\\ 
		(c)  \hspace{3.5 cm}  (d)
	\end{center}
	\caption{Four views of decimated 3D point cloud from laser scanner and fitted ground truth planes corresponding to the four walls, ceiling and floor. Figure (a) shows an assonometric view of the  room while Figure (b), (c) and (d) show top and two lateral views.}
	\label{fig:GTplanes}
	\vspace{-.2cm}
\end{figure}
\begin{figure}[tbhp]
	\begin{center}
		\includegraphics[width=.9\linewidth]{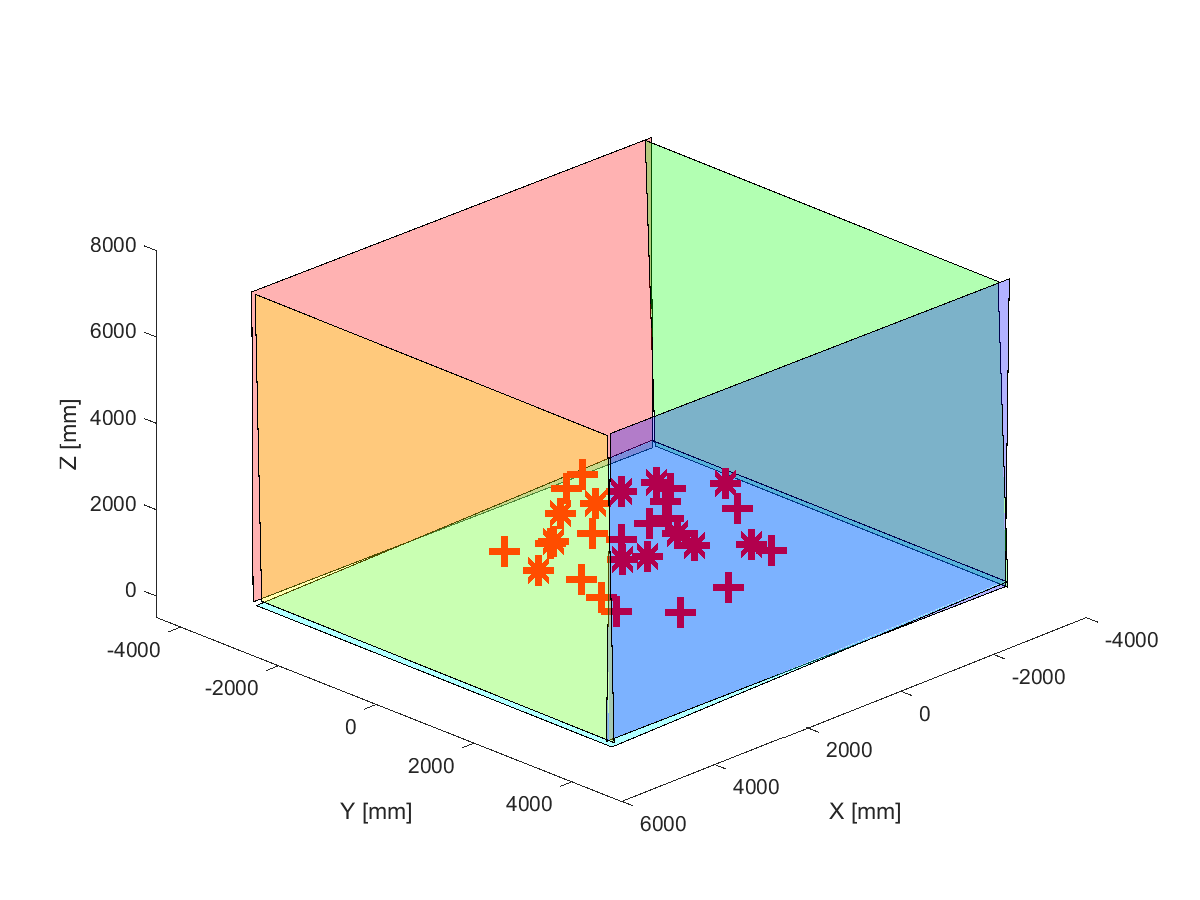}  
	\end{center}
	\vspace{-.35cm}
	\caption{Overall ground truth data: 3D displacement of microphones (stars) and audio sources (crosses) from motion capture system, and planar surfaces (fitted from point cloud yielded by the laser scanner).}
	\label{fig:GTcomplete}
	\vspace{-.2cm}
\end{figure}
In the great part of works concerning room reconstruction, ground truth data are manually acquired, e.g. measuring distances by a tape, thus limiting their precision. On the contrary we made an effort in acquiring a precise ground truth position of microphones, sources and room boundaries. In particular, we employed a Leica C10 laser scanner to acquire a high resolution point cloud of the environment geometry, as displayed in Figure \ref{fig:laser_scanner}. The 3D point cloud was manually divided into six subsets, each one corresponding to the points belonging to the four walls, the floor and the ceiling respectively and a linear regression was performed on each subset to estimate the plane parameters.
In Fig. \ref{fig:GTplanes} the six estimated planes together with a decimated point cloud is displayed in four views. Notice the vaulted ceiling and the niches corresponding to the windows that present relevant differences from the assumed piecewise planar model.

To estimate sources and microphones positions we used a VICON motion capture system with 8 BONITA cameras. Measurements were made by placing the cameras at the room borders having a $4m \times 4m \times 2m$ recording area and by attaching a marker to each microphone and speaker center. 
The registration between 3D scan and motion capture coordinate systems was achieved thanks to $6$ additional markers placed on the floor, visible by both the VICON and the 3D scanner device.   

The overall ground truth data, enclosing microphones and sources 3D positions and planes fitted to the 3D point cloud are displayed in Fig. \ref{fig:GTcomplete}. Notice that the relative displacement of microphones and sources with respect to planes limits the possibility to apply heuristics to solve the labelling problem. In fact, depending on the source position and on the microphone considered, the order of arrival of peaks corresponding to virtual sources is not constant, especially for echoes due to lateral walls. 
%
%
\begin{figure}[tbhp]
	\begin{center}
		\includegraphics[width=1\linewidth]{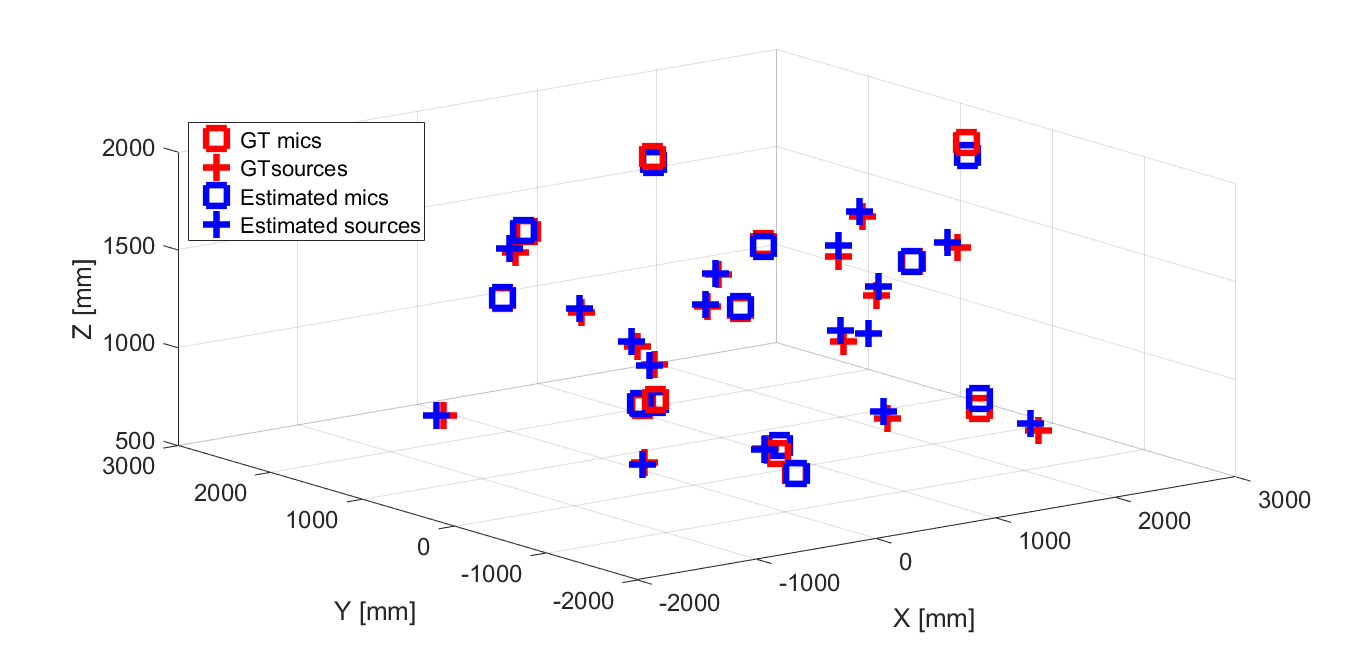} 
	\end{center}
	\caption{Ground truth and estimated microphone and source 3D displacements.}
	\label{fig:msrec}
\end{figure}


\subsection{Results and Discussion}

In Fig. \ref{fig:msrec} the 3D positions of ground truth (red) and estimated (blue) microphones (squares) and sources (crosses) are displayed. It can be seen that the overall displacement is correctly reconstructed without outliers in the 3D position of microphones and sources. The mean and standard deviation of the reconstruction error for sources and microphones is  computed as the Euclidean distance between ground truth and estimation as reported in Table \ref{lambda}. Reconstruction accuracy is slightly better for microphones (average error of $22mm$) than for sources (average error of $31mm$). Considering the actual size of microphones and sources, respectively about $1cm$ and $4cm$ of diameter for the active surface, we can conclude that a reasonable accuracy is achieved despite the deviation from the employed model assuming omnmidirectional, point-like sources and microphones.

\begin{table}[bth]
	\centering{
		\begin{tabular}{|c| c | c |}
			\hline
			& \textbf{sources}   & \textbf{microphones} \\
			\hline			
			mean  & 59 mm & 31 mm \\
			\hline
			std & 31 mm & 22 mm \\
			\hline
		\end{tabular}
	}
	\vspace{.2cm}
	\caption{Mean and standard deviation errors for microphones and sources 3D displacements.}
	\label{lambda}
\end{table}

\begin{figure}[tbhp]
	\begin{center}
		\includegraphics[width=\linewidth]{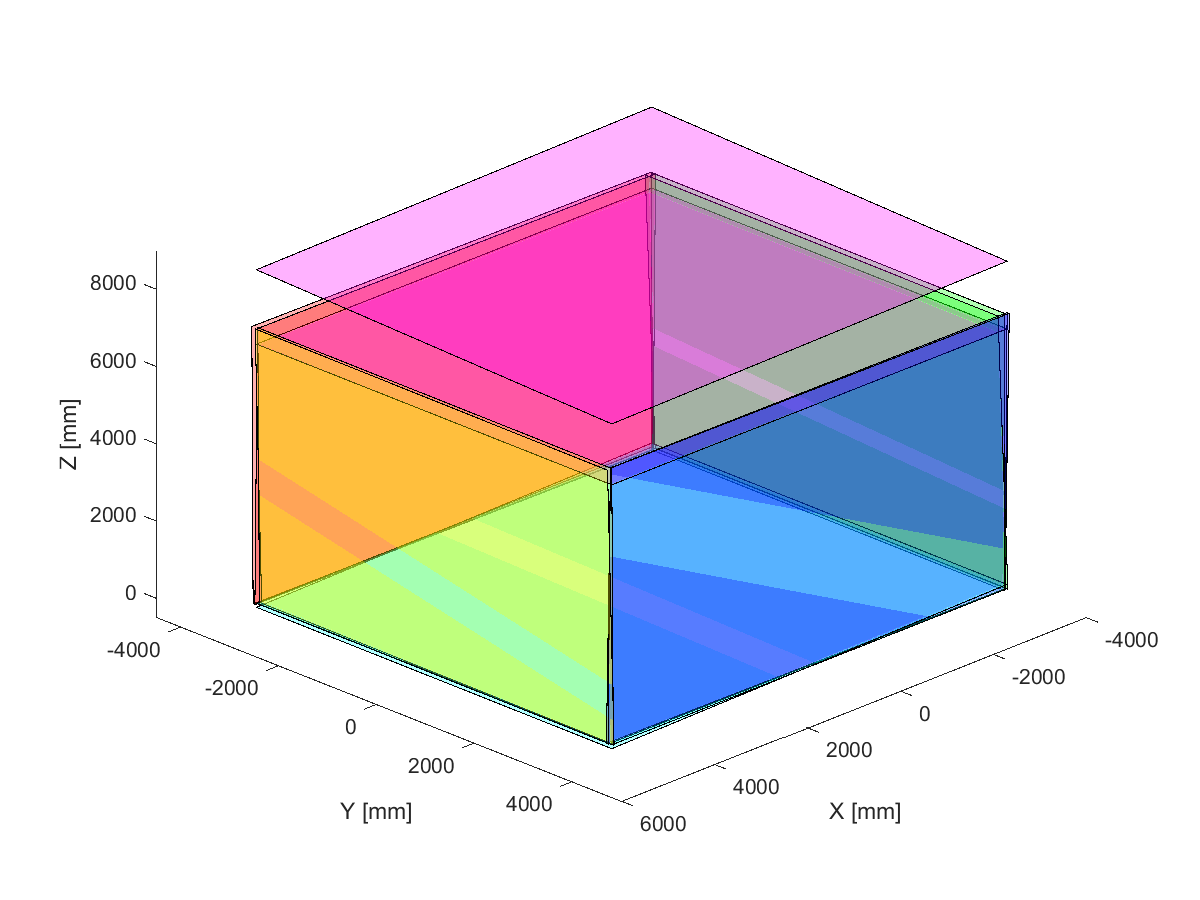}  
	\end{center}
	\vspace{-.35cm}
	\caption{Ground truth and estimated planes with the refined solution corresponding to the four walls, floor and ceiling.}
	\label{fig:GTestplanes}
	\vspace{-.2cm}
\end{figure}

\begin{figure}[bth]
	\begin{center}
		\includegraphics[width=\linewidth]{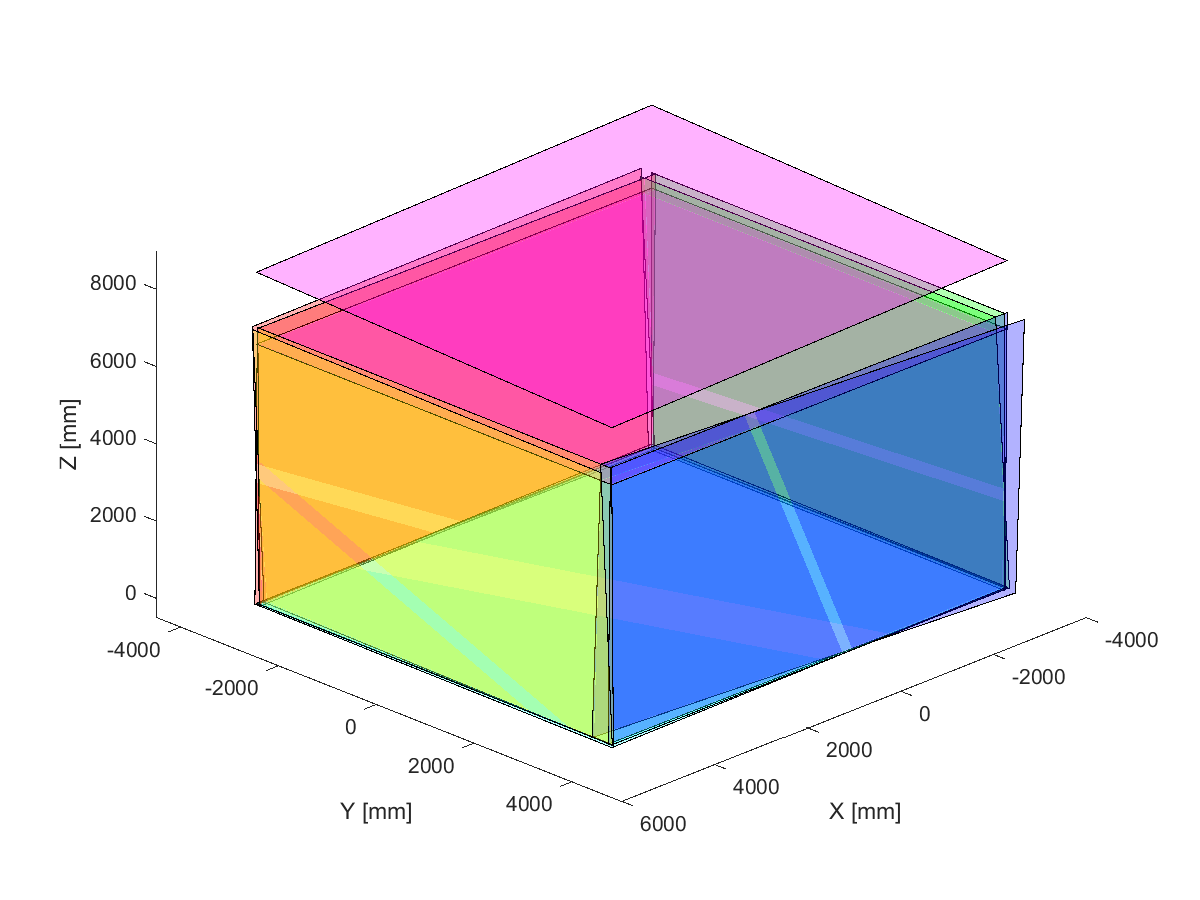}  
	\end{center}
	\vspace{-.35cm}
	\caption{Ground truth and estimated planes with the first guess solution corresponding to the four walls, floor and ceiling.}
	\label{fig:GTestplanes_first_guess}
	\vspace{-.2cm}
\end{figure}

\begin{table}[tbhp!]
	\centering{
		\begin{tabular}{| c | c |c|}
			\hline
			& \textbf{angle error} &  \textbf{distance error} \\
			\hline			
			wall1   & 0.89 deg  &  22 mm\\
			\hline
			wall2   &  1.72 deg  & 34 mm\\
			\hline
			wall3   &  0.63 deg & 62 mm\\
			\hline
			wall4   &  0.43 deg & 4 mm\\
			\hline
			floor   &  1.40 deg & 1 mm\\
			\hline
			ceiling     &  2.98 deg & 1824 mm\\
			\hline
		\end{tabular}
	}
	\vspace{.2cm}
	\caption{Mismatch between ground truth and estimated planes according to the refined solution. Angle error is the solid angle between the normals of planes and distance error is the absolute difference in distances of planes form the room center.}
	\label{lambda1}
\end{table} 

\begin{table}[thbp!]
	\centering{
		\begin{tabular}{| c | c |c|}
			\hline
			& \textbf{angle error} &  \textbf{distance error} \\
			\hline			
			wall1   & 1.93 deg  &  66 mm\\ 
			\hline
			wall2   &  2.41 deg  & 68 mm\\
			\hline
			wall3   &  1.99 deg & 98 mm\\
			\hline
			wall4   &  3.82 deg & 1 mm\\
			\hline
			floor   &  0.82 deg & 2 mm\\
			\hline
			ceiling     &  3.57 deg & 1795 mm\\
			\hline
		\end{tabular}
	}
	\vspace{.2cm}
	\caption{Mismatch between ground truth and estimated planes according to the first guess solution. Angle error is the solid angle between the normals of planes and distance error is the absolute difference in distances of planes form the room center.}
	\label{lambda2}
\end{table}

Concerning the room reconstruction in Fig. \ref{fig:GTestplanes_first_guess}, the six estimated planes are overlapped to the ground truth ones showing a remarkable accuracy for the four walls and the floor and a qualitatively correct estimation of the ceiling position. This difference is easily explainable considering the vaulted ceiling of the room that makes the ceiling height not well defined and causes a relevant different from the assumed piecewise planar model of the room surfaces. The estimation accuracy for the surfaces is reported in Table \ref{lambda1}: excluding the ceiling, the angle between normal vectors ranges from $0.43$ degrees to $1.72$ degrees error while the difference in normal vector lengths ranges from $1mm$ to $62mm$. To make evident the effect of final global refinement with respect to the solution based on the grid search, described in Sections III.D and III.C respectively, the surface reconstruction based on grid search is displayed in Fig. \ref{fig:GTestplanes_first_guess} and numerical errors are reported in Table \ref{lambda2}. Results are qualitatively correct but significantly less precise than the refined solution, with an angle error ranging from $0.82$ degrees to $3.82$ degrees and a distance error ranging from $1mm$ to $98mm$ (excluding the ceiling).


\section{Conclusions and future work}

We presented a full pipeline for room geometry reconstruction from sound where the only assumption made is about room convexity and known signal in transmission. In such scenario, the reconstruction achieves remarkable results in an affordable computational time, even for unfavourable SNR values. Moreover we presented an evaluation on a real dataset with accurate ground truth. This data will be made available to the community\footnote{Dataset with ground truth can be freely downloaded at: {www.iit.it/datasets/vgm-3d-room-reconstruction-dataset}} so providing a reference for evaluating these methods in unconstrained scenarios. Future work will be dedicated to the treatment of non-covex rooms (e.g. \cite{occal2014source,Bedri:2014}) or using unknown natural signals (e.g. exploiting the methods for blind RIR estimation  \cite{Crocco:DelBue:2015,Crocco:DelBue:2016}). 

%
%

\section*{Acknowledgment}

The authors would like to thank Carlos~Beltr\'an-Gonz\'alez and Cosimo Rubino for helping with the real experimental setup deployment and acquisition.


\bibliographystyle{IEEEtran}
\bibliography{biblio_journal}



%
%
%
%
%



\end{document}